\documentclass[a4paper]{article}

\usepackage{etoolbox}
\usepackage[utf8x]{inputenc}
\usepackage{color,colortbl}
\usepackage{bbding}
\usepackage{multirow}
\usepackage{lscape}
\usepackage{array,graphicx}
\usepackage{textcomp}
\usepackage{epstopdf}
\usepackage{threeparttable}
\usepackage{caption}
\usepackage{subcaption}
\usepackage{bbm}
\usepackage{nicefrac}
\usepackage{afterpage}
\usepackage{amsmath,amsfonts,amssymb,amsthm}
\usepackage{newpxtext} 

\usepackage{authblk}
\usepackage{hyperref}

\usepackage{float}
\usepackage[margin=1in]{geometry}

\usepackage{tocloft}

\patchcmd{\thebibliography}{\chapter*}{\section*}{}{}

\definecolor{Gray}{gray}{0.95}

\definecolor{Gray}{gray}{0.95}

\begin{document}

\title{Identifying partners at sea on contrasting fisheries around the world}

\author[1,2,*]{\small Roc\'io Joo}
\author[3]{\small Nicolas Bez}
\author[4]{\small Marie-Pierre Etienne}
\author[5]{\small Pablo Marin}
\author[6]{\small Nicolas Goascoz}
\author[2]{\small J\'er\^ome Roux}
\author[2]{\small St\'ephanie Mah\'evas}

 \affil[1]{\footnotesize Department of Wildlife Ecology and Conservation, Fort Lauderdale Research and Education Center, University of Florida, Fort Lauderdale, FL, USA}
\affil[2]{\footnotesize IFREMER, Ecologie et Mod\`eles pour l'Halieutique, BP 21105, 44311 Nantes Cedex 03, France}
\affil[3]{\footnotesize MARBEC, Univ Montpellier, IRD, Ifremer, CNRS, S\`ete, France}
\affil[4]{\footnotesize Univ Rennes, Agrocampus Ouest, CNRS, IRMAR - UMR 6625, F-35000 Rennes, France}
\affil[5]{\footnotesize IFREMER, Laboratoire Ressources Halieutiques de Port en Bessin, BP 32, 14520 Port en Bessin, France}
\affil[6]{\footnotesize Instituto del Mar del Per\'u (IMARPE), Chucuito, Callao, Peru}
\affil[*]{Corresponding author: Roc\'io Joo, rocio.joo@ufl.edu}

\date{}

\maketitle 

Running title: Dyadic joint movement in world fisheries 

\newpage

\begin{abstract}

Here we present an approach to identify partners at sea based on fishing track analysis, and describe this behaviour in six different fleets: 1) pelagic pair trawlers, 2) large bottom otter trawlers, 3) small bottom otter trawlers, 4) mid-water otter trawlers, all operating in the North-East Atlantic Ocean, 5) anchovy purse-seiners in the South-East Pacific Ocean, and 6) tuna purse-seiners in the Western Indian Ocean.
This type of behaviour is known to exist within pelagic pair trawlers. Since these vessels need to be in pairs for their fishing operations, in practice some of them decide to move together throughout their whole fishing trips, and others for only a segment of their trips. 
To identify partners at sea, we used a heuristic approach based on joint-movement metrics and Gaussian mixture models. The models were first fitted on the pelagic pair trawlers and then used on the other fleets. From all of these fisheries, only the tuna purse-seiners did not present partners at sea. 
We then analysed the connections at the scale of vessels and identified exclusive partners. 
This work shows that there are collective tactics at least at a pairwise level in diverse fisheries in the world. 

\end{abstract}


%
 
\section{Introduction}

Understanding fisher spatial behaviour contributes to the development of effective spatial management policies. The increasing availability of georeferenced data from sources like Automatic Identification System (AIS; \cite{Robards2016}) and Vessel Monitoring System (VMS; \cite{Hinz2013}) has enabled a proliferation of studies that characterise fisher spatial behaviour (e.g. \cite{Bertrand2005,Joo2014}), propose movement models (e.g. \cite{Vermard2010,Walker2010,Joo2013,Gloaguen2015a}), account for it in stock management models (e.g. \cite{Vigier2018}) and discuss management measures based on it (e.g. \cite{Gerritsen2012,Holmes2011}). While individual (or independent) movement of fishers has been extensively studied by means of trajectory data, the collective behaviour of fishermen has been rather neglected. Fishers are social individuals that may develop collaboration or competing strategies (e.g. \cite{Horta2012,Hancock1995}). The characterisation of their collective behaviour could provide valuable inputs that would make movement models more accurate and management measures more effective \cite[]{Salas2004,Rijnsdorp2011,Gezelius2007}, as they would rely in more realistic scenarios.  

Collective behaviour can be produced at large or small group scales, and may be reflected in a variety of movement patterns. Here we focus on dyadic or pairwise joint movement behaviour, and more specifically, partners at sea, defined by a couple of fishing vessels that move together during their time at sea. 
An extensive review and comparison of metrics for assessing dyadic joint movement \cite[]{Joo2018} showed that the metrics varied in their sensitivity to three aspects of joint movement: proximity, coordination in direction and coordination in speed.
Partners at sea should show coordinated and proximal joint movement. To account for all of these aspects, we chose one metric for each of the three dimensions of joint movement, from the ones recommended in \cite{Joo2018}, to characterise the dyadic movement of fishing vessels. 

Strong partnership at sea was expected to be found in pelagic pair trawlers: since they need to be in pairs at least during each fishing operation, they are likely to be paring throughout their entire fishing trips. While this has not been systematically studied, this pattern has been observed. 
For that reason, we focused first on a pelagic pair trawlers dataset in the North-East Atlantic Ocean to learn about metric patterns that could be revealing partners-at-sea behaviour.
We analysed their VMS data to identify dyads or potential candidates for partners at sea and computed the three joint movement metrics for each dyad. Then, we fitted a three-component Gaussian mixture model (GMM) to distinguish three groups of dyads sharing the same types of behaviour.
One of these components was expected to correspond to partners-at-sea patterns.
After characterising at-sea partnership in this fleet, we used the fitted model to identify partners at sea in several other fisheries in the world: bottom and mid-water otter trawlers in the North-East Atlantic Ocean, anchovy purse-seiners in the South-East Pacific Ocean, and tuna purse-seiners in the Western Indian Ocean. 
We showed that this type of behaviour is not exclusive to pelagic pair trawlers, and discuss possible implications of this behaviour in terms of fishing strategies. 
Perspectives opened by this work for further research in collective spatial behaviour are also discussed. 

\section{Materials and Methods}

\subsection{Fishing vessels trajectory data}

In this section, the VMS data and fishing trip characteristics of the analysed fleets are briefly described. These are: $1$) French pelagic pair trawlers, $2$) French large bottom otter trawlers, $3$) French small bottom otter trawlers, $4$) French mid-water otter trawlers, all operating in the North-East Atlantic Ocean,  $5$) French tuna purse-seiners in the Western Indian Ocean, and $6$) Peruvian anchovy purse-seiners in the South-East Pacific Ocean.

For the French fleets, the use of VMS started to be legislated and mandatory in the European Union since 2000. 
In practice, records are transmitted at $\sim$ 1 h intervals. In the North-East Atlantic Ocean, we analysed VMS data from fishing trips performed between 2012 and 2013 within the English Channel and the Celtic Sea, while in the Indian Ocean, we analysed fishing trips from 2011 to 2013. In Peru, industrial purse-seiners are also legally obliged to use VMS tracking devices since 2000, transmitting their positions at $\sim$ 1 h intervals, but since 2015, VMS positions are recorded each $10$ minutes. We focus on Peruvian fishing trips during a specific fishing season in 2016.


\subsubsection{French pelagic pair trawlers} 
\label{PTM}

A pelagic pair trawl is a gear defined by one trawl towed in midwater by two vessels to target pelagic fish. Thus, vessels of the pelagic pair trawler fleet remain close performing almost synchronous movements while operating the trawl. However, they do not need to move together throughout their whole fishing trips, especially when steaming, using single trawls or exploring the sea individually looking for shoals. These vessels can spend part of their fishing trips on individual activities, even targetting other fish that do not require pair trawling.
Most of the pair-trawler fishing trips in the dataset were performed by relatively large vessels (18-24 m; $\sim 80\%$), and they last $\sim 99h$ on average, according to fisher logbooks.

\subsubsection{French large and small bottom otter trawlers} 
\label{LOTB-SOTB}

The bottom otter trawl gear is a trawl towed by a single vessel; these vessels target bottom and demersal species. Vessels performing bottom otter trawl fishing trips had a large variability in their sizes: from 10 to 40 m. The duration of the trips were proportionally related to the size of the vessels: larger vessels performed longer trips and generally offshore. Since, for this type of gear, the spatial behaviour from smaller vessels differ from that of larger vessels (e.g. the trips are not only shorter but also closer to the coast), we separated bottom otter trawlers into two groups: one with vessels smaller than 12 m or performing trips of less than 20 h (we assume that in very short trips even large vessels act like the small ones), and another one with vessels larger than 12 m or performing trips of larger duration; vessels with these characteristics are considered as composing the small otter trawl and large otter trawl fishing fleets, respectively. The average duration of fishing trips for both fleets were $\sim$ 16 and $\sim$ 105 hours, respectively, according to fisher logbooks.

\subsubsection{French mid-water otter trawlers} 
\label{OTM}

A mid-water otter trawl gear is also operated by an individual vessel, where otter boards hold the mouth of the net open. As the vessels in the pair trawler fleet, mid-water otter trawlers target pelagic fish mostly. As with bottom trawlers, vessels performing mid-water trawling trips had sizes ranging from 10 to 40 m; larger vessels exist (e.g. 90 m long targeting blue whiting) but were not found in this dataset. However, the spatial behaviour of these vessels was not conditioned by their size, so they were not separated by size. The average duration of a fishing trip was $\sim$ 31 hours (fisher logbooks). Since fishing with mid-water or bottom otter trawls does not require pair-work, if it exists, it would reflect a strategic/tactical choice. 

\subsubsection{French tuna purse-seiners}

The fleet is composed of ten to twenty vessels operating in the Indian Ocean and the size of the purse seiners is typically of sixty meters.
Tuna purse-seiners' fishing trips usually last several tens of days. 
The time windows targeted in the present study (2011-2013) followed a harsh period of strong security issues induced by piracy attacks in the Indian Ocean. During the second half of 2009, it became mandatory for fishing vessels operating in the Indian Ocean to fish in pairs before some military protection were enforced. However, some vessels could have decided to continue moving more or less in pairs as a precautionary approach. Since tuna purse-seiners perform long fishing trips, we did not expect vessels to move together throughout their whole fishing trips, but rather over some shorter opportunist periods of time, eventually changing partners.

\subsubsection{Peruvian anchovy purse-seiners}

The ten-minutes frequency of data recording is particularly suiting for monitoring the anchovy (\textit{Engraulis ringens}) industrial fishery, where fishing trips usually last less than 24 hours (a median of $17$ hours for the analysed data), since fish tends to distribute close to the coast in dense patches \cite[]{Bertrand2008,Joo2014}. In this fishery, vessel size is measured in terms of its hold capacity, which varies from $32.5$ MT to $900$ MT, with a median at $\sim 100$ MT. We used data from the first fishing season of 2016 (39 days between June and July). Though the race for fish stopped in 2009 (the total allowable catch was replaced by an individual vessel quota system; \cite{Aranda2009}), the high abundance of anchovy, the eagerness to save fuel oil and the habit of performing very short fishing trips, make it common for vessels to go to the same fishing zones or to follow each other as a fishing tactic. Thus here as well, we expected to find some patterns of joint movement, although not perfectly synchronous or remaining close to each other all the time.

\subsection{Methods}

\begin{figure}[ht!]
	\centering%
	\includegraphics[scale=0.45]{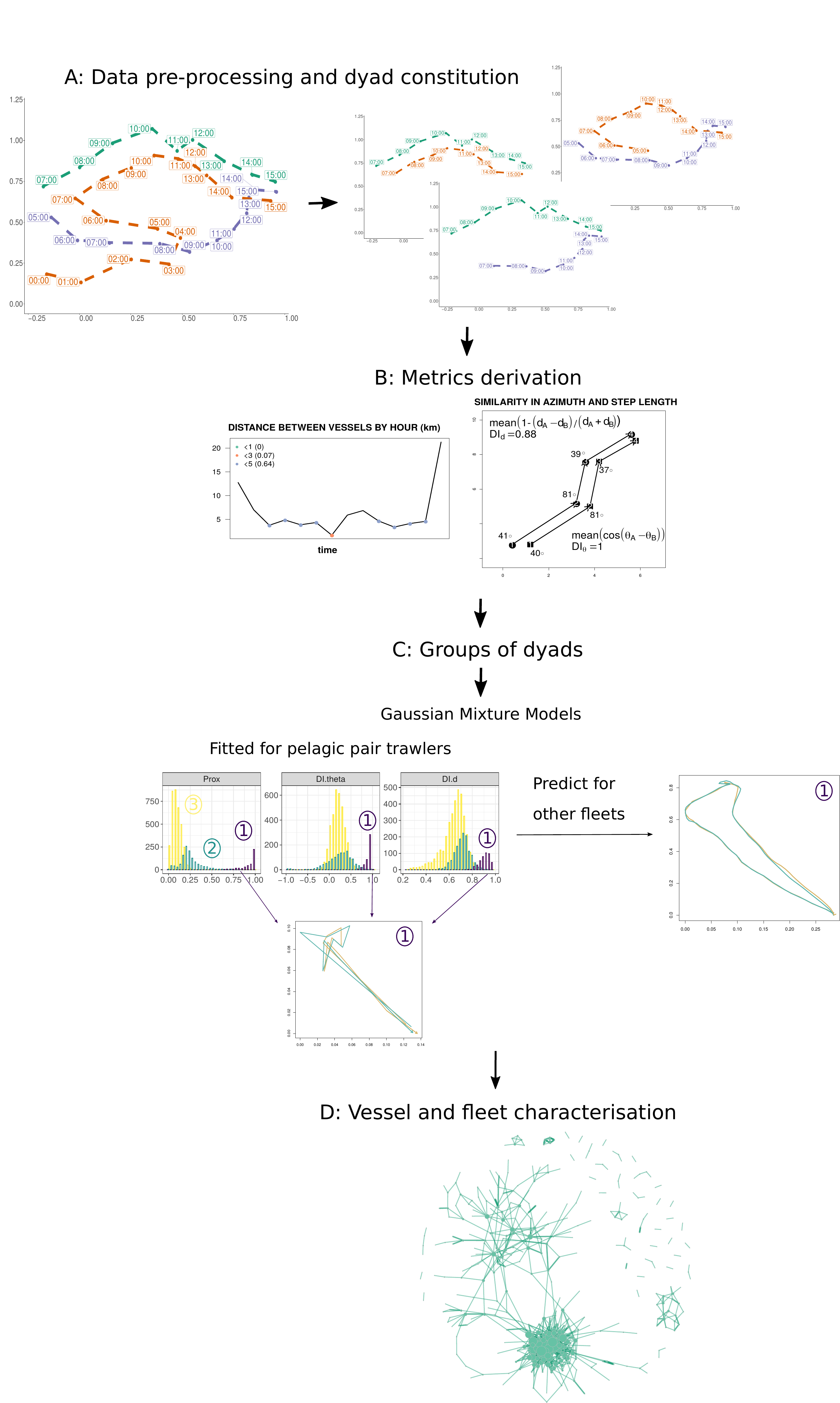}
	\caption{Stages of dyadic joint movement analysis by fleet. A: Data pre-processing and dyad constitution (couple of track segments at sea at the same time; see in the example that three vessel tracks result in three dyads that depend on their time at sea together). B: metrics derivation for each dyad. C: cluster analysis of the dyads, mainly fitting a Gaussian mixture model to pelagic pair trawlers and using the model to identify partners at sea in other fleets. D: vessel and fleet characterisation based on the clusters.}
	\label{FigMethods}
\end{figure}

Identifying partners at sea basically consists of 1) data pre-processing and dyad constitution (i.e. the VMS data was first cleaned and interpolated, and then dyadic segments of trajectories were identified); 2) joint-movement metrics derivation for each dyad; 3) identification of clusters of dyadic joint movement --and particularly partners at sea-- via GMMs; and 4) characterisation of partnership at vessel and fleet scales (Fig. \ref{FigMethods}). All the analyses were performed in R \cite[]{R2015}.

\subsubsection{Data Pre-processing}

Fishing trips where at least one pair of consecutive records were lagged by more than three hours were removed ($\approx 9\%$ of the total number of fishing trips). Then, since location records had irregular time steps, we interpolated tracks to obtain regular steps and simultaneous VMS positions (i.e. fixes) from vessels at sea. A linear interpolation method was used, since we considered than a one-hour-step linear interpolation for records that were already separated by one hour on average (for trawlers and tuna purse-seiners data), or a 10-minute-step linear interpolation for records that were already separated by 10 minutes on average (for anchovy purse-seiners data) should not be too far from the `real' locations.

From the (interpolated) fixes, we derived motion variables such as displacement (distance between consecutive fixes) and absolute angle (between the direction of the x-axis and the locations at consecutive fixes). The adehabitat package in R \cite[]{adehabitatLT} was used to compute those metrics. 

We then formed the dyads that would be candidates for partners at sea. Dyads were defined as the concomitant parts of two vessel tracks crossing each other at least once during their fishing trips. 
We considered that, to `cross each other', vessels had to be at a proximity of $<5$ km at least once for all fleets, except tuna purse-seiners. The latter have a greater range of motion and do not get so close; for them, the distance threshold was set to 60 km. If both vessels departed from port and then arrived to port at the same time, the dyad was to be composed of the two tracks of their whole fishing trips; if not, the dyad would have been composed by track segments of their fishing trips corresponding to moments when both vessels were at sea (see graphical example in part A of Fig. \ref{FigMethods}). To keep only dyads with segments that were long enough for the analysis, an arbitrary 10-hour threshold was set for all trawlers and anchovy purse-seiner fleets. Tuna purse-seiners performed longer trips, so the 10th percentile was used as their threshold. The number of vessels, dyads and the median duration of a dyad are shown in Table \ref{Table:Dyads}.

\begin{table}[ht!]
	\caption{Statistics per fleet of number of vessels, number of dyads, their duration (median in hours), the $\delta$ threshold for Prox, and the frequency of record transmission. The first three statistics are also displayed for each cluster.} 
	\begin{center}
		\begin{tabular}{cc|cccc}
			\hline
			\hline
			\rule{0pt}{11pt}
			&  & \multirow{2}{*}{Total} & \multicolumn{3}{c}{Clusters} \\
			&  & & $1$ & $2$ & $3$ \\
			\hline 
			\rule{0pt}{11pt}
			Pelagic &  \multirow{2}{*}{Vessels} & \multirow{2}{*}{$59$} & $56$ & $57$ & $58$ \\
			pair & & & $(94.9\%)$ & $(96.6\%)$ & $(98.3\%)$ \\
			trawlers & \multirow{2}{*}{Dyads} & \multirow{2}{*}{$6457$} & $495$ & $1681$ & $4281$ \\
			($\delta= 5km$, & & & $(7.7\%)$ & $(26.0\%)$ & $(66.3\%)$ \\
			$\Delta t=1h$) & Duration & $87$ & $74$ & $68$ & $97$ \\
			\hline 
			\rule{0pt}{11pt}
			Large & \multirow{2}{*}{Vessels} & \multirow{2}{*}{$266$} & $38$ & $254$ & $261$ \\
			bottom & & & $(14.3\%)$ & $(95.5\%)$ & $(98.1\%)$ \\
			otter trawlers & \multirow{2}{*}{Dyads} & \multirow{2}{*}{$54478$} & $312$ & $16205$ & $37961$ \\
			($\delta= 5km$, & & & $(0.6\%)$ & $(29.8\%)$ & $(69.7\%)$ \\			
			$\Delta t=1h$) & Duration & $65$ & $60$ & $47$ & $73$ \\
			\hline 
			\rule{0pt}{11pt}
			Small & \multirow{2}{*}{Vessels} & \multirow{2}{*}{$202$} & $52$ & $185$ & $183$ \\
			bottom & & & $(25.7\%)$ & $(91.6\%)$ & $(90.6\%)$ \\
			otter trawlers & \multirow{2}{*}{Dyads} & \multirow{2}{*}{$17300$} & $93$ & $7051$ & $10156$ \\
			($\delta = 3km$, & & & $(0.5\%)$ & $(40.8\%)$ & $(58.7\%)$ \\		
			$\Delta t = 1h$) & Duration & $12$ & $12$ & $12$ & $12$ \\	
			\hline 
			\rule{0pt}{11pt}
			Mid-water & \multirow{2}{*}{Vessels} & \multirow{2}{*}{$70$} & $4$ & $56$ & $65$ \\
			otter & & & $(5.7\%)$ & $(80.0\%)$ & $(92.9\%)$ \\
			trawlers & \multirow{2}{*}{Dyads} & \multirow{2}{*}{$844$} & $3$ & $409$ & $432$ \\
			($\delta= 3km$, & & & $(0.4\%)$ & $(48.5\%)$ & $(51.2\%)$ \\			
			$\Delta t=1h$)	& Duration & $12$ & $11$ & $12$ & $12$ \\	
			\hline 
			\rule{0pt}{11pt}
			Anchovy & \multirow{2}{*}{Vessels} & \multirow{2}{*}{$757$} & $327$ & $756$ & $756$ \\
			purse- & & & $(43.2\%)$ & $(99.9\%)$ & $(99.9\%)$ \\
			seiners & \multirow{2}{*}{Dyads} & \multirow{2}{*}{$572804$} & $568$ & $168284$ & $403952$ \\
			($\delta=3 km$, & & & $(0.1\%)$ & $(29.4\%)$ & $(70.5\%)$ \\		
			$\Delta t= 10 min$) & Duration & $17$ & $16$ & $16$ & $17$ \\	
			\hline 
			\rule{0pt}{11pt}
			Tuna & \multirow{2}{*}{Vessels} & \multirow{2}{*}{$15$} & $0$ & $15$ & $15$ \\
			purse- & & & $(0.0\%)$ & $(100.0\%)$ & $(100.0\%)$ \\
			seiners & \multirow{2}{*}{Dyads} & \multirow{2}{*}{$1523$} & $0$ & $39$ & $1484$ \\
			($\delta= 10km$, & & & $(0.0\%)$ & $(2.6\%)$ & $(97.4\%)$ \\			
			$\Delta t= 1h$) & Duration & $357$ & $-$ & $224$ & $362$ \\	
			\hline
			\hline
		\end{tabular}
	\end{center}
	\label{Table:Dyads}
\end{table} 

\subsubsection{Joint movement metrics} 

The review made by \cite{Joo2018} defined three dimensions of joint movement: proximity (closeness in space-time), coordination in direction and coordination in speed. The article evaluated ten metrics used in the literature to assess joint movement and showed that some metrics were either redundant or inaccurate for characterising joint movement, some others were better suited to assess proximity, and others were more sensitive to coordination. Based on that work, we chose three metrics that were positively evaluated and that -- together -- account for the different aspects of joint movement: 1) the proximity index (proximity), 2) dynamic interaction in displacement (coordination in speed, and in displacement when time steps are regularly spaced), and 3) dynamic interaction in direction (coordination in direction). 

The proximity index (Prox) is defined as the proportion of simultaneous fixes that are spatially close. To define closeness, we needed to fix a distance threshold $\delta$. For pair trawlers, it is expected that at the very moment of fishing, vessels working together are separated by less than 1 km from each other. When they were not fishing, they could still move together but not necessarily at $<1$km. Thus, a $5$km threshold was used for this fleet. 
We also used a 5km threshold for large bottom otter trawlers to get comparable results to those of pair trawlers. Anchovy purse-seiners, mid-water, and small bottom otter trawlers usually perform short and coastal fishing trips, meaning that vessels would not necessarily move together as a strategy, but could sometimes coincide in places due to their short coastal movements. For that reason, we chose a smaller threshold, 3km, for those three fleets. For tuna purse-seiners, we chose 10km, as it is roughly the limit of visual detection of neighbouring vessels. 

The calculation of the other two metrics did not require an \textit{ad hoc} parametrization as for Prox. The dynamic interaction in direction ($DI_\theta$) and in displacement ($DI_{d}$) measured similarity in direction and speed/displacement, respectively, between simultaneous fixes (i.e. records of locations) in a dyad. The mathematical definition of each metric is shown in Table \ref{IndTable}.

\begin{table}[ht!]
	\caption{Joint movement metrics}
	\begin{center}
		\begin{tabular}{lll}
			\hline
			\hline
			\rule{0pt}{11pt}
			Metric & Range & Interpretation for joint movement \\ 
			\hline 
			\rule{0pt}{11pt}
			$Prox = \left(\displaystyle\sum_{t=1}^{T}\mathbbm{1}{\{d_t^{A,B} < \delta \} }\right)/T$ & $$\quad [0,1]$$ & From always distant (0) to \\
			& & always close (1) \\
			$DI_d = \left(\displaystyle\sum_{t=1}^{T-1} \left[1 - \left(\frac{\mid d_{t,t+1}^{A}-d_{t,t+1}^{B}\mid}{d_{t,t+1}^{A}+d_{t,t+1}^{B}}\right)^\beta\right]\right)/(T-1)$ & $$\quad [0,1]$$ & From non-cohesive (0) to cohesive (1) \\
			& & movement in displacement \\
			$DI_\theta = \left(\displaystyle\sum_{t=1}^{T-1} \cos(\theta_{A_t} - \theta_{B_t})\right)/(T-1)$ & $$\quad [-1,1]$$ & From opposite (-1) to cohesive (1) \\
			& & movement in azimuth \\
			\hline
			\hline
		\end{tabular}
	\end{center}
	\begin{tablenotes}
		\small
		\item \textit{Note:}  $A$, $B$: vessels in the dyad; $T$: number of fixes in the dyad; $d_t^{A,B}$: distance in km between vessels $A$ and $B$ at $t$-th fixes; $\mathbbm{1}{ \{\}}$: index function; 
		$\delta$: distance threshold; 
		$d_{t,t+1}^{A}$ (resp. $d_{t,t+1}^{B}$): displacement of $A$ (resp. $B$) in km between fixes $t$ and $t+1$; $\beta$ is a scaling parameter for which we assume to take the default value of 1 \cite[]{Long2013,Joo2018};
		$\theta_{A_t}$ (resp. $\theta_{B_t}$): heading of vessel $A$ (resp. $B$) at time $t$. 
	\end{tablenotes}
	\label{IndTable}
\end{table} 

\subsubsection{Identification of partners at see with Gaussian mixture models}

Partner identification was addressed through a probabilistic clustering approach using GMMs \cite[]{biernacki2006model}. In this approach, each dyad $i$ was characterised by its three dimensional metrics $X_i=(Prox_i, {DI_d}_i, {DI_\theta}_i)$ which were assumed to be a realisation of a three-dimensional normal distribution. The mean vector and the variance matrix of this distribution depended on the unknown cluster $Z_i$ to which the dyad $i$ belonged. 
Given a fixed number of clusters $(G)$ and the three metrics, there were three elements to estimate for each cluster $g$ $(g=1,...,G)$: a three-dimensional mean $(\mu_g)$, a $3 \times 3$ covariance matrix $(\Sigma_g)$, and the proportion of the cluster in the observed dyad population $(\pi_g)$.

In this set-up, the probability density function of given metric values $x_i$ of a dyad $i$ ($\phi(x_i)$) can be expressed as:


$$\phi(x_i)=\sum_{g=1}^{G} \pi_g f_g(x_i,\mu_g,\Sigma_g)$$ where $\pi_g = P(Z_i = g)$ and $ f_g(x_i,\mu_g,\Sigma_g)$ is a three-dimensional Gaussian density function.


The probability of being in cluster $g$ for each dyad $i$ given the observed metrics, $P(Z_i = g \vert X_i = x_i)$, also called posterior probability, was obtained as a by-product of the global estimation of the model and is expressed as follows:

$$P(Z_i=g \vert X_i=x_i) = \frac{ \pi_g f_g(x_i,\widehat{\mu}_g,\widehat{\Sigma}_g) }{\displaystyle \sum_{k=1}^{G}\pi_k f_g(x_i,\widehat{\mu}_k,\widehat{\Sigma}_k) }, $$
where $\widehat{\mu}_g$ and $\widehat{\Sigma}_g$ stand respectively for the estimated mean in cluster $g$ and the corresponding estimated covariance matrix.

In GMMs, the total number of clusters are chosen according to either statistical selection criteria (mostly likelihood-based) or case-study goals.
A three-component GMM structure, i.e. $G=3$, was chosen in order to obtain higher discrepancies between two extreme dyadic-behaviour clusters by allowing to have a cluster in between corresponding to an intermediate behaviour.

Each covariance matrix $\Sigma_g$ can be expressed as the product of different components which specify its orientation, shape and volume (see  \cite{biernacki2006model}). We chose a general GMM structure of 3 dyadic-behaviour clusters allowing for the volume, orientation and shape of the clusters to differ from one another, called Gaussian\_pk\_Lk\_Ck in  \cite{biernacki2006model}.

The GMMs were fitted to the pelagic pair trawlers dataset, composed of $6457$ dyads. 
Parameter estimation was achieved via the iterative EM algorithm.
Because EM is known to be sensitive to initial conditions \cite[]{dempster1977maximum}, we fitted 30 different GMMs and kept the one that minimised the integrated complete likelihood criterion, using the Rmixmod package \cite[]{Rmixmod} and based on \cite{biernacki2006model}. From the fitted model, henceforth denoted by $GMM_{pair trawlers}$, we obtained the posterior probability $P(Z_i=g \mid X_i=x_i)$ of each dyad $i$ to belong to each cluster $g$ given the metric values $x_i$. 
We considered that a dyad was classified as part of the cluster $g$ that maximised the posterior probability $P(Z_i=g \mid X_i=x_i)$. 
The level of mixture between pairs of clusters in the final model was quantified as the overlapping volume between the tri-Gaussian distributions of each cluster. This index ranges between 0 (no mixture) and 1 full (mixing). High levels of mixture would indicate that the clusters are difficult to distinguish from each other, making the classification poorly relevant. 

For each cluster, we computed a global average of the Z-scores (i.e. centred and scaled transformation) of their $(Prox_i, {DI_d}_i, {DI_\theta}_i)$-features, and ordered them accordingly. Based on the definitions of the metrics \cite[]{Joo2018}, the cluster with the highest average was associated to partners-at-sea behaviour. 

The GMM fitted on pelagic pair trawlers ($GMM_{pair trawlers}$) was then used on each of the other fleets to classify their dyads, into the three identified groups. For each dyad $i$ of the other fleets, we computed $P(Z_i=g \mid X_i=x_i)$ for $g=\{1,2,3\}$ and assigned the dyad to the most plausible cluster.

\subsubsection{Vessel and fleet characterisation} 

We focused on the first group of each fleet, i.e. partners at sea. Their relative importance in the fleets were represented by the proportions of vessels and dyads involved in the cluster.
For each fleet, the social relationships between vessels that engaged at least once in partners at sea behaviour were visually represented as a social network \cite[]{Scott1988,Jacoby2016}. The elements of the sociomatrix of the network, i.e. adjacency matrix, represented the number of partner-at-sea dyads between the vessels ---that had at least one dyad in the cluster. 
The Fruchterman and Reingold algorithm was chosen to draw the graph. It positions the nodes of the graph in the space so that all edges are more or less equal length and there are as few crossing edges as possible, aiming at an aesthetic representation \cite[]{Fruchterman1991}. The igraph package was used for this purpose \cite[]{igraph2006R}. 

We identified which and how many vessels were exclusive, i.e. only formed partners at sea with one vessel throughout the whole period of study. In the adjacency matrix this corresponded to the rows with 0 everywhere except once. To assess how exclusive were partnerships at the fleet level, a loyalty index was defined as the proportion of vessels that showed exclusiveness in partnership.

All the R codes for partner-at-sea identification via GMMs and vessel and fleet characterisation are available at \url{https://rociojoo.github.io/partners-at-sea/} (doi:10.5281/zenodo.4016377).

\section{Results}

\subsection{Pelagic pair trawlers}

\begin{table}[ht!]
\caption{$\pi$ and $\mu$ estimates of $GMM_{pair trawlers}$}
\begin{center}
	\begin{tabular}{ll|ccc}
		\hline
		\hline
		\rule{0pt}{11pt}
		& & \multicolumn{3}{c}{Clusters} \\
		& & 1 & 2 & 3 \\
		\hline
		\multicolumn{2}{l|}{$\pi$} & $0.08$ & $0.33$ & $0.59$ \\
		\multirow{3}{*}{$\mu$} & Prox & $0.94$ & $0.20$ & $0.09$ \\
		& $DI_{\theta}$ & $0.93$ & $0.23$ & $0.18$ \\
		& $DI_d$ & $0.92$ & $0.70$ & $0.63$ \\
		\hline
		\hline
	\end{tabular}
\end{center}
\label{Table:GMM-mean}
\end{table} 
	
\begin{table}[ht!]
	\caption{$\Sigma$ estimates of $GMM_{pair trawlers}$}
	\begin{center}
		\begin{tabular}{l|ccc|ccc|ccc}
			\hline
			\hline
			\rule{0pt}{11pt}
			& \multicolumn{9}{c}{Clusters} \\
			& \multicolumn{3}{c|}{1} & \multicolumn{3}{c|}{2} & \multicolumn{3}{c}{3} \\
			\hline
			& Prox & $DI_{\theta}$ & $DI_d$ & Prox & $DI_{\theta}$ & $DI_d$ & Prox & $DI_{\theta}$ & $DI_d$ \\
			Prox & $0.007$ & $0.003$ & $0.001$ & $0.016$ & $0.015$ & $0.002$ & $0.003$ & $0.003$ & $0.001$ \\
			$DI_{\theta}$ & $0.003$ & $0.005$ & $0.002$ & $0.015$ & $0.063$ & $0.006$ & $0.003$ & $0.024$ & $0.007$ \\
			$DI_d$ & $0.001$ & $0.002$ & $0.002$ & $0.002$ & $0.006$ & $0.004$ & $0.001$ & $0.007$ & $0.010$ \\
			\hline
			\hline
		\end{tabular}
	\end{center}
	\label{Table:GMM-variance}
\end{table} 

\begin{figure}[ht!]
	
	\begin{minipage}[b]{.5\linewidth}
		\centering%
		\includegraphics[scale=0.5]{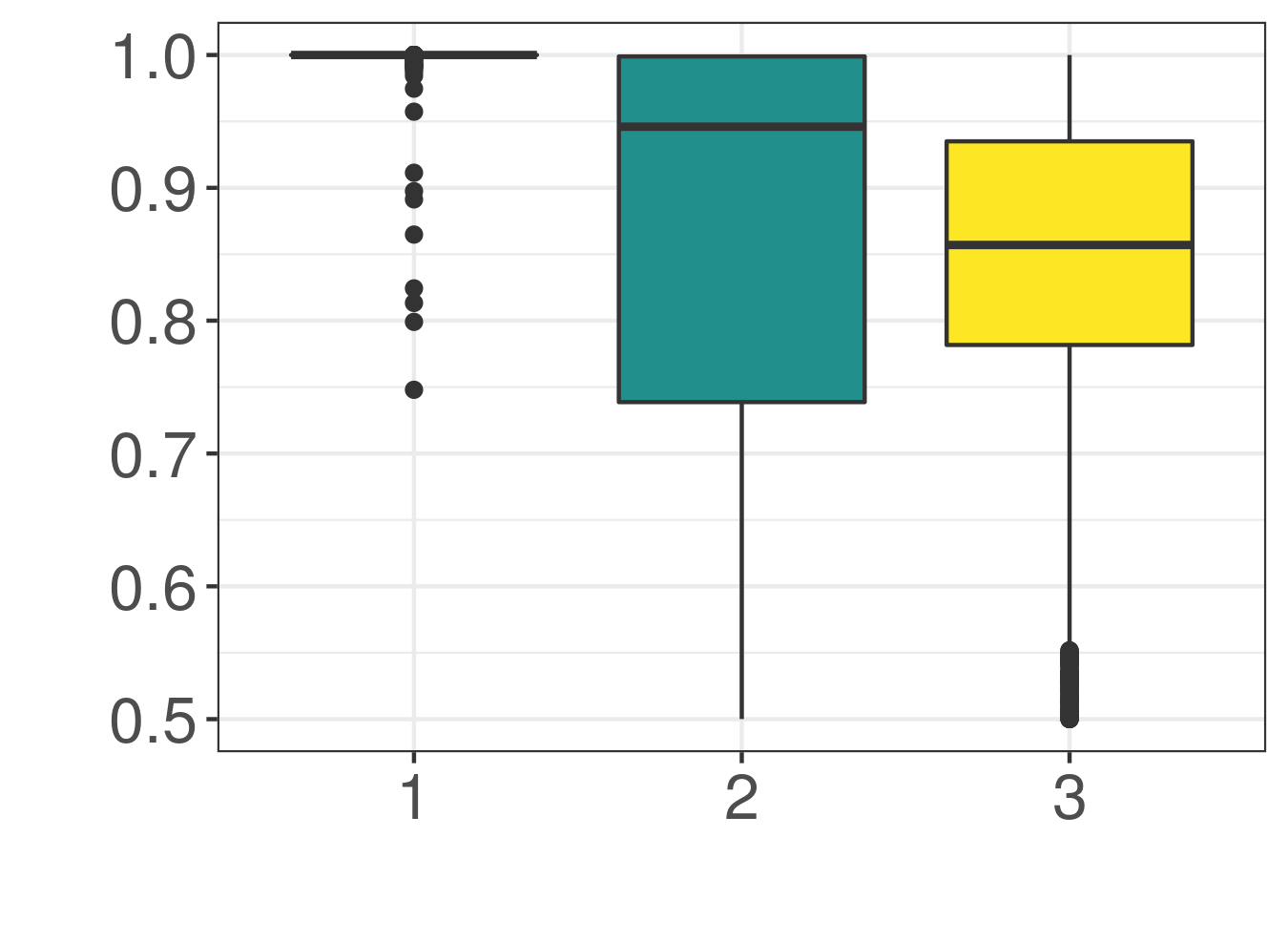}
		\subcaption{Pelagic pair trawlers}
		\label{PTMpca}
	\end{minipage}
	\hfill
	\begin{minipage}[b]{.5\linewidth}
		\centering%
		\includegraphics[scale=0.5]{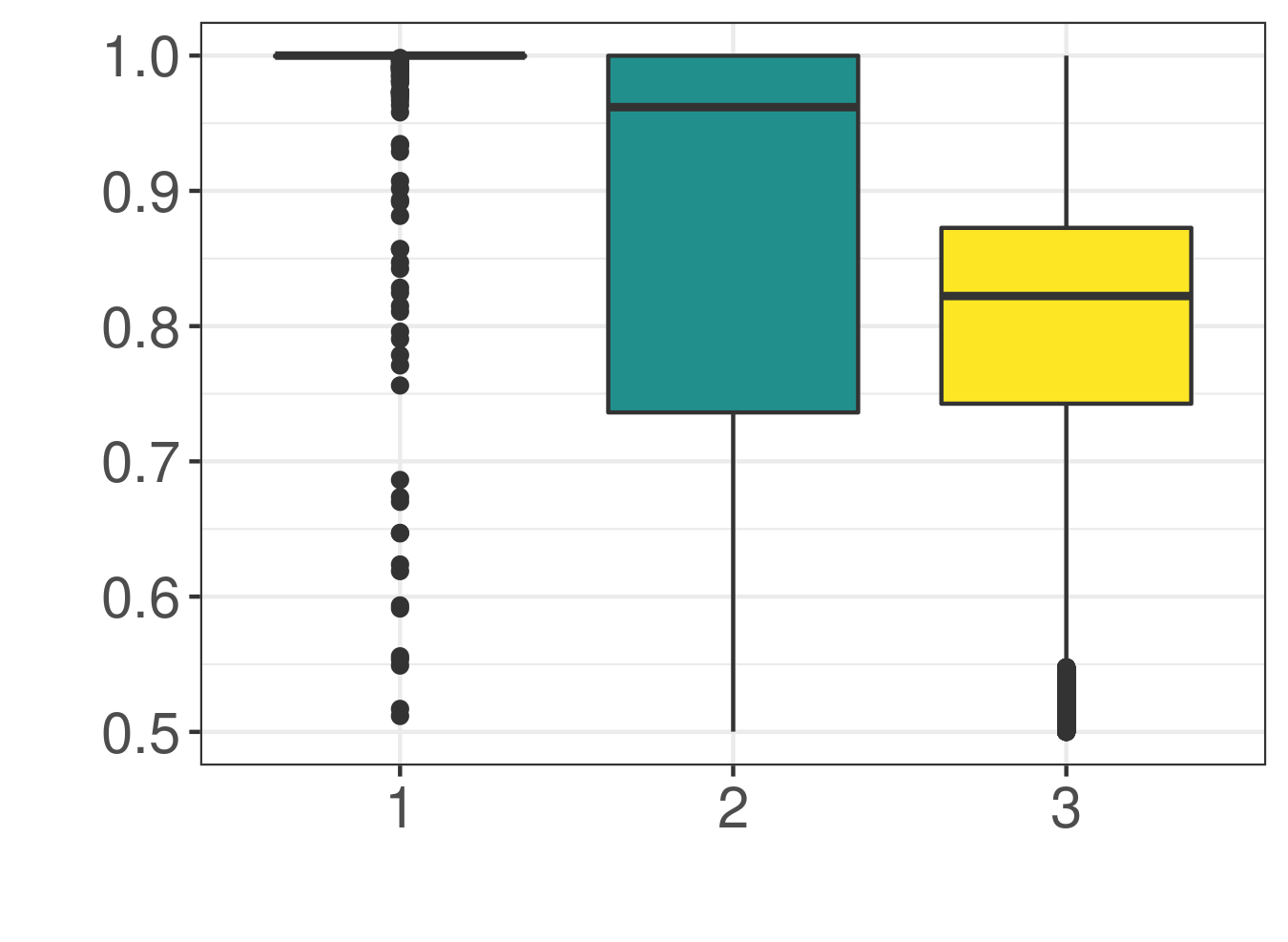}
		\subcaption{Large bottom otter trawlers}
		\label{LOTBpca}
	\end{minipage}\\
\vspace{0.3cm} \\
	\begin{minipage}[b]{.5\linewidth}
		\centering%
		\includegraphics[scale=0.5]{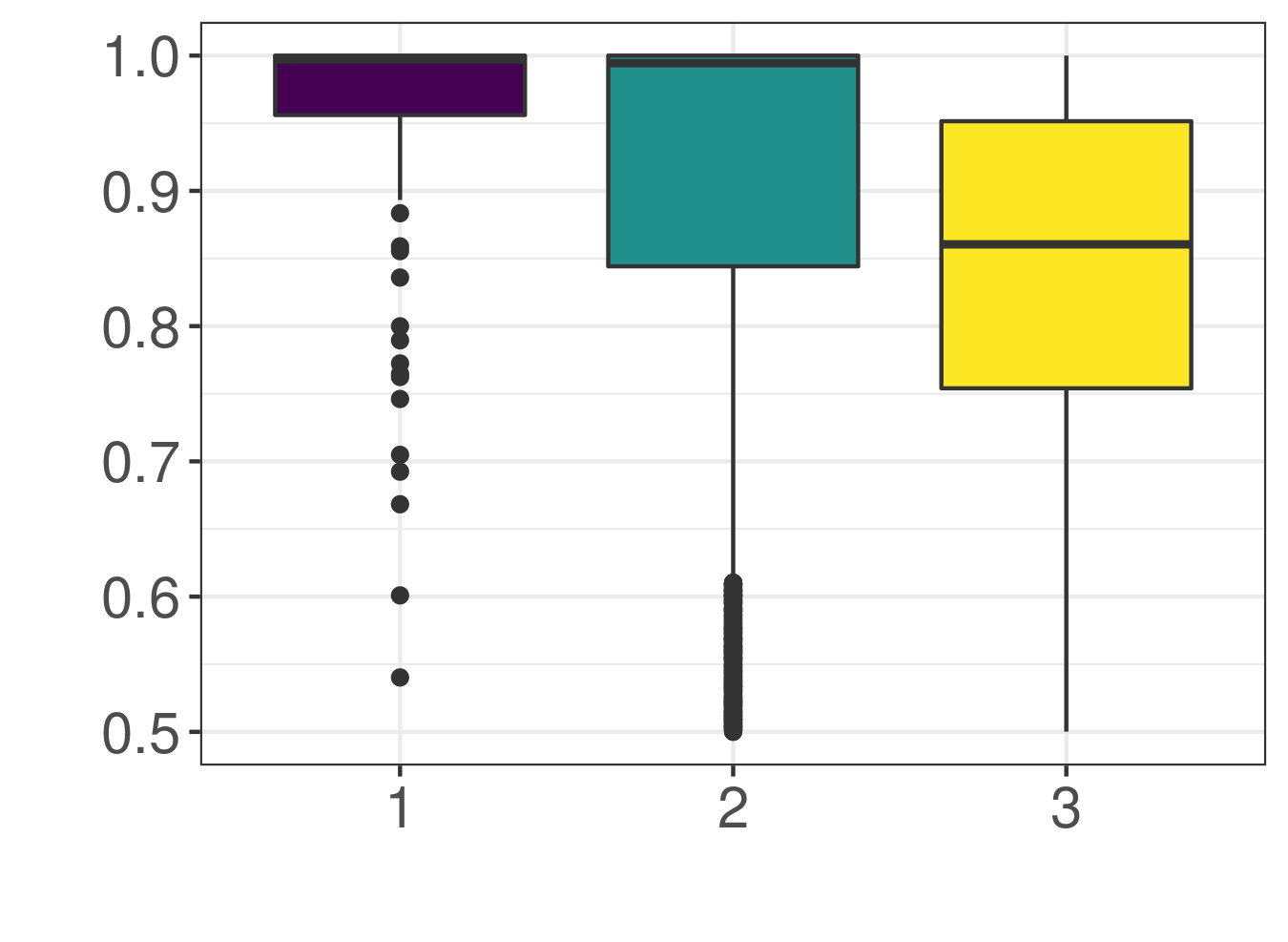}
		\subcaption{Small bottom otter trawlers}
		\label{SOTBpca}
	\end{minipage}
	\hfill
	\begin{minipage}[b]{.5\linewidth}
		\centering%
		\includegraphics[scale=0.5]{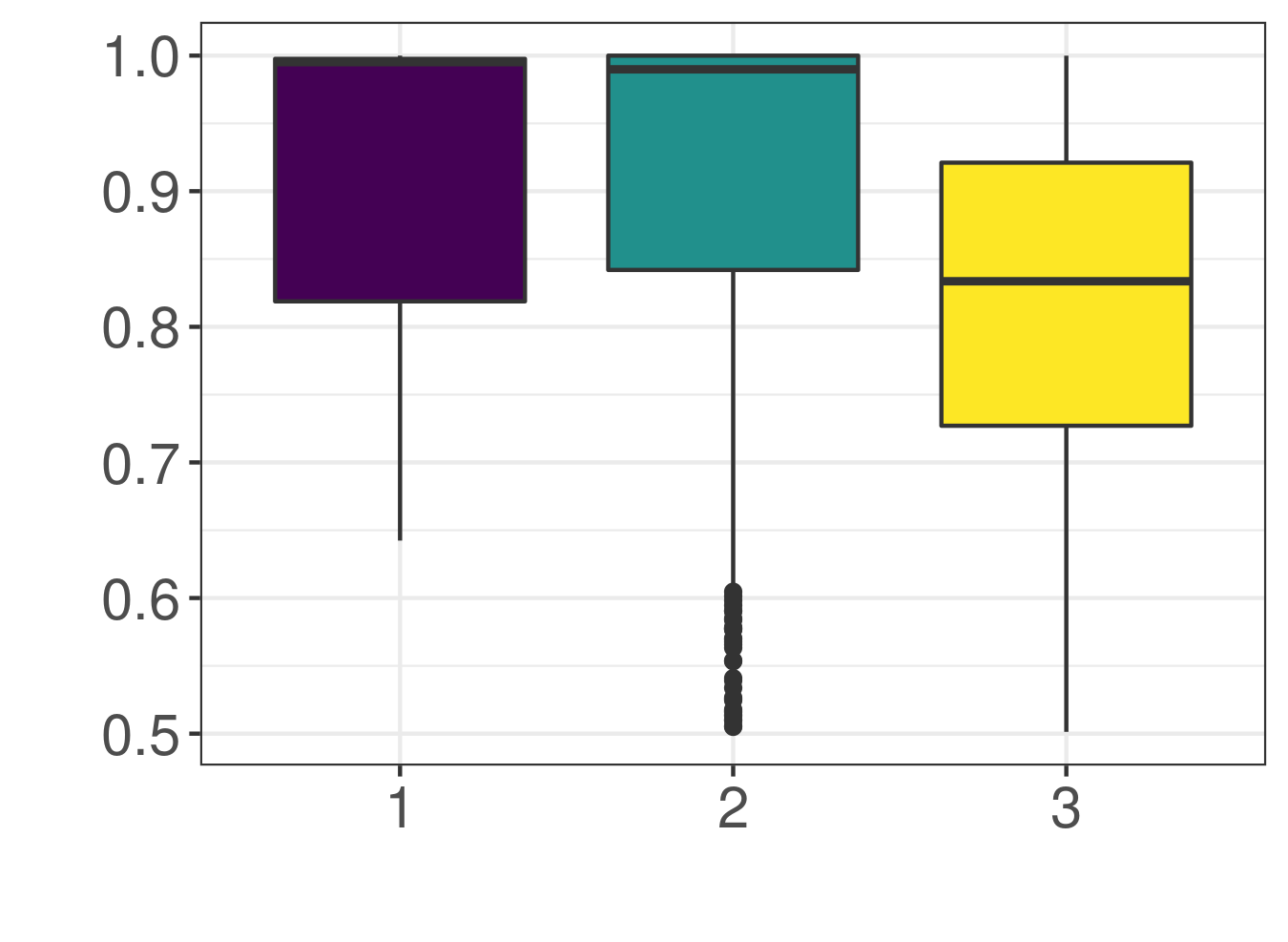}
		\subcaption{Mid-water otter trawlers}		
		\label{OTMpca}	
	\end{minipage}
\vspace{0.3cm} \\
	\begin{minipage}[b]{.5\linewidth}
		\centering%
		\includegraphics[scale=0.5]{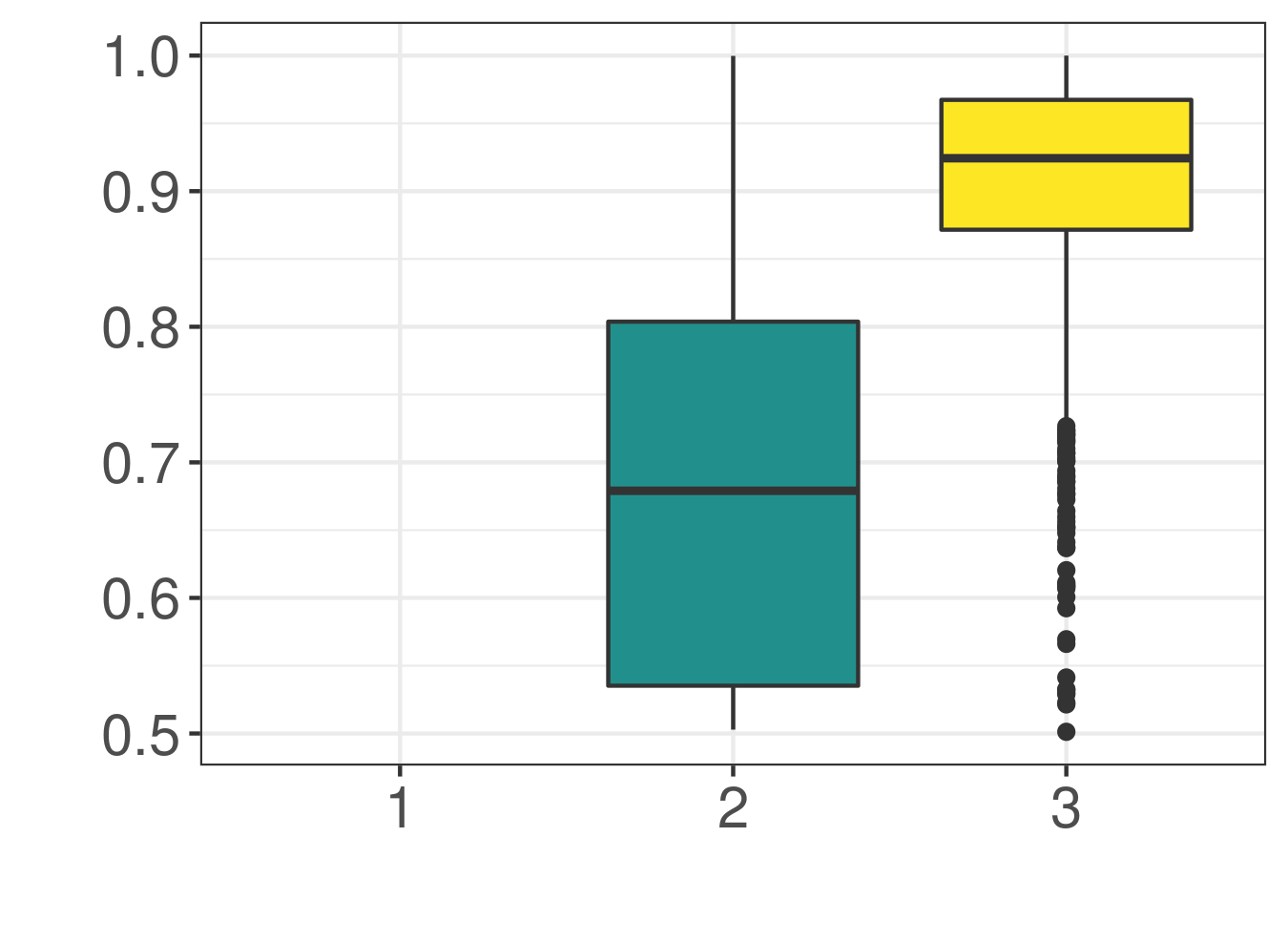}
		\subcaption{Tuna purse-seiners}
		\label{TPSpca}	
	\end{minipage}
	\hfill
	\begin{minipage}[b]{.5\linewidth}
		\centering%
		\includegraphics[scale=0.5]{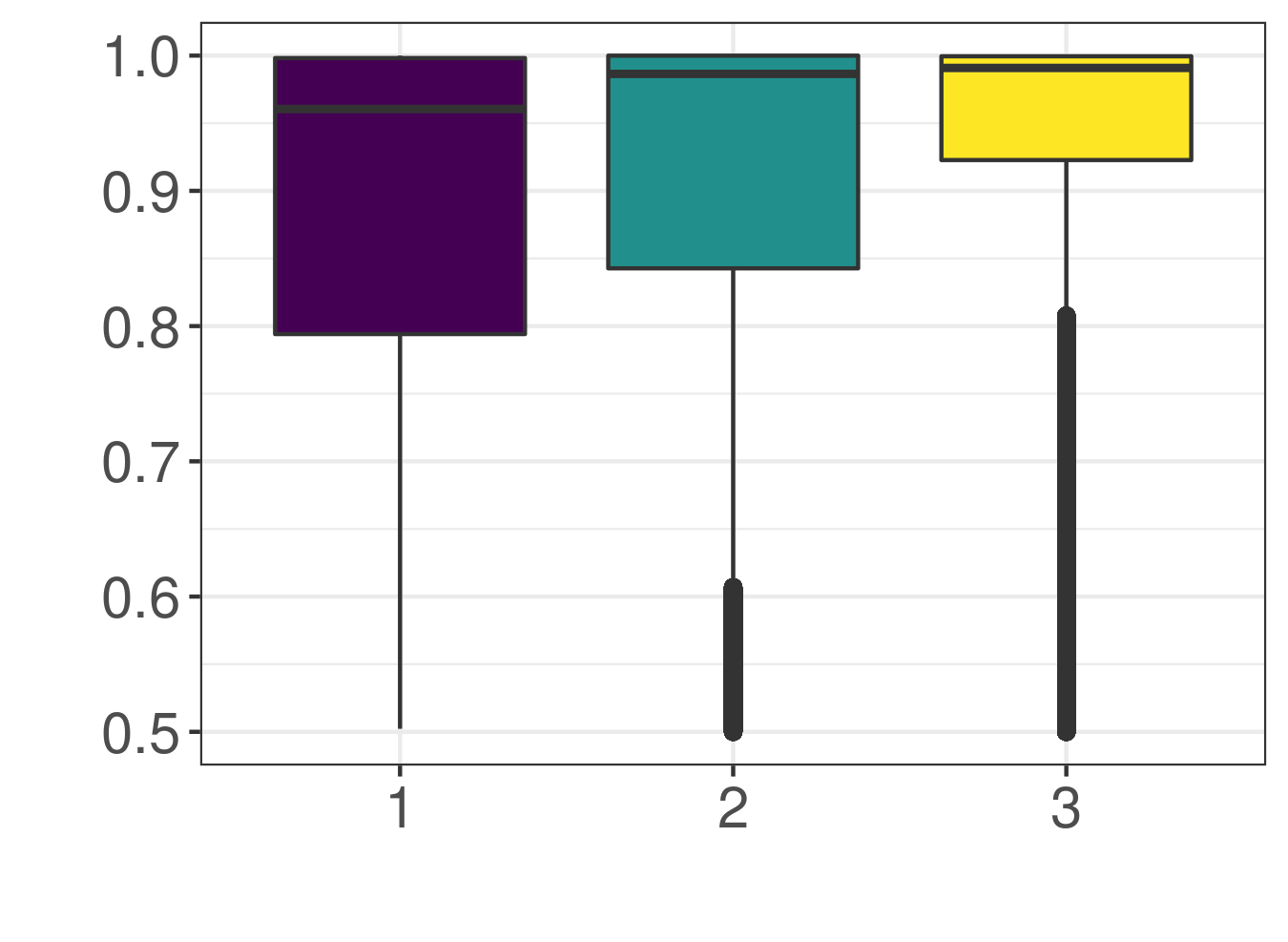}
		\subcaption{Anchovy purse-seiners}
		\label{APSpca}
	\end{minipage}			
	\caption{Boxplots of the posterior probabilities $P(Z_i = g| X_i = x_i)$ of each dyad $i$ classified in each group.}
	\label{Fig:proba}
\end{figure}

After pre-processing, $6457$ dyads were classified with GMMs. The estimated parameters are shown in Tables \ref{Table:GMM-mean} and \ref{Table:GMM-variance}. The covariance between features ($\Sigma$) was not negligible, which supports the joint use of metrics that evaluate different aspects of dyadic movement. There was little overlap between cluster 1 and the other two: $1.9 \times 10^{-3}$ and $3.7 \times 10^{-10}$, between clusters 1 and 2, and 1 and 3, respectively. There was higher overlap ($0.32$) between clusters 2 and 3. Moreover, most dyads were classified based on high values of their posteriors ($1.00$, $0.95$, and $0.86$ as median posteriors for each group, respectively; Fig. \ref{Fig:proba}), and all of them above $0.5$.

\begin{figure}[ht!]
\centering%
		\includegraphics[scale=0.55]{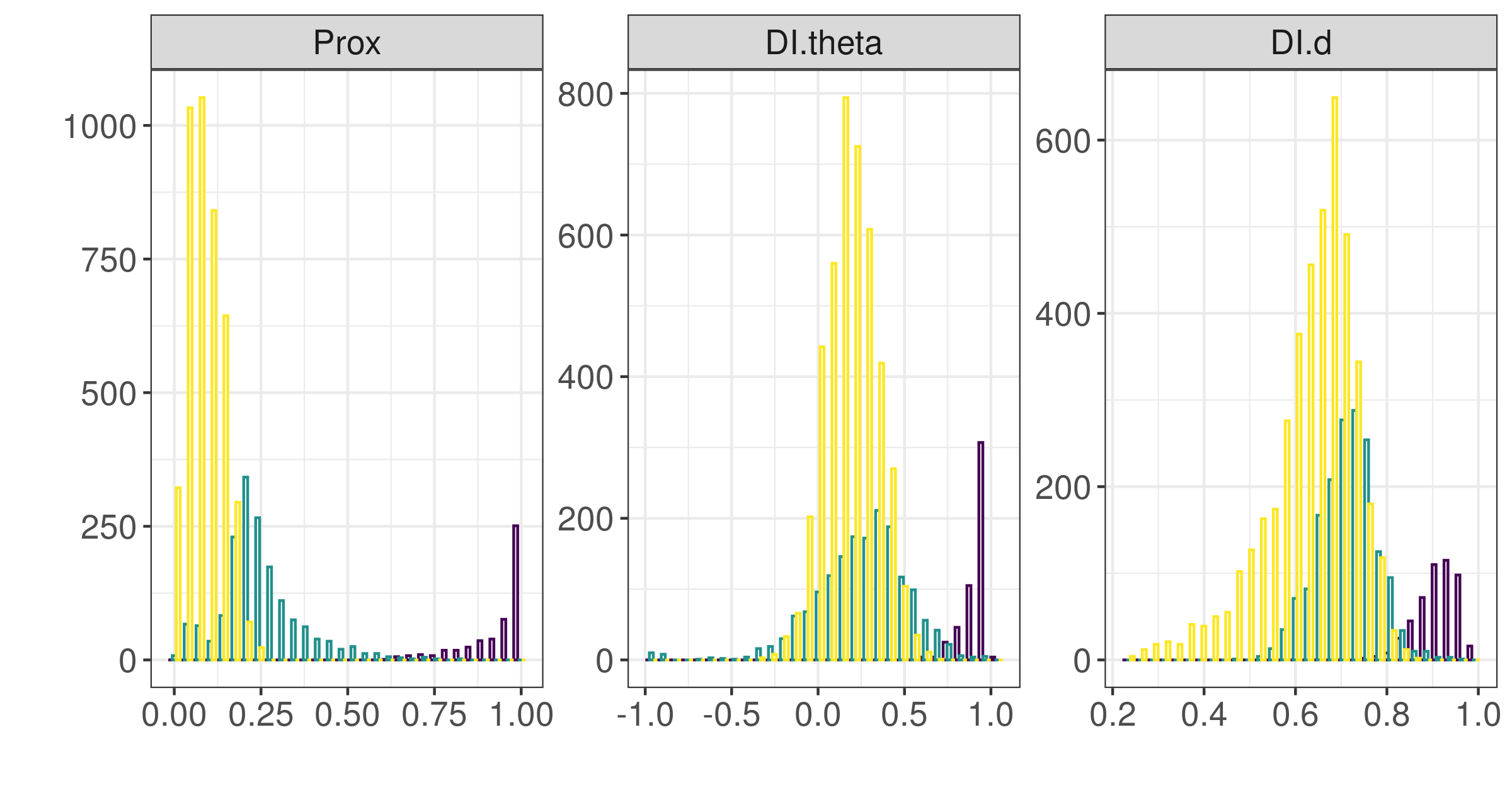} \\	
	\caption{Histograms of the joint movement metrics for the three clusters (in purple, green and yellow) for pelagic pair trawlers. It should be noted that only $DI_\theta$ ranges from -1 to 1, while Prox and $DI_d$ take values from 0 to 1.}
	\label{Fig:Clusters}
\end{figure}

\begin{figure}[ht!]	
	\begin{minipage}[b]{.5\linewidth}
		\centering
		\includegraphics[scale=0.4]{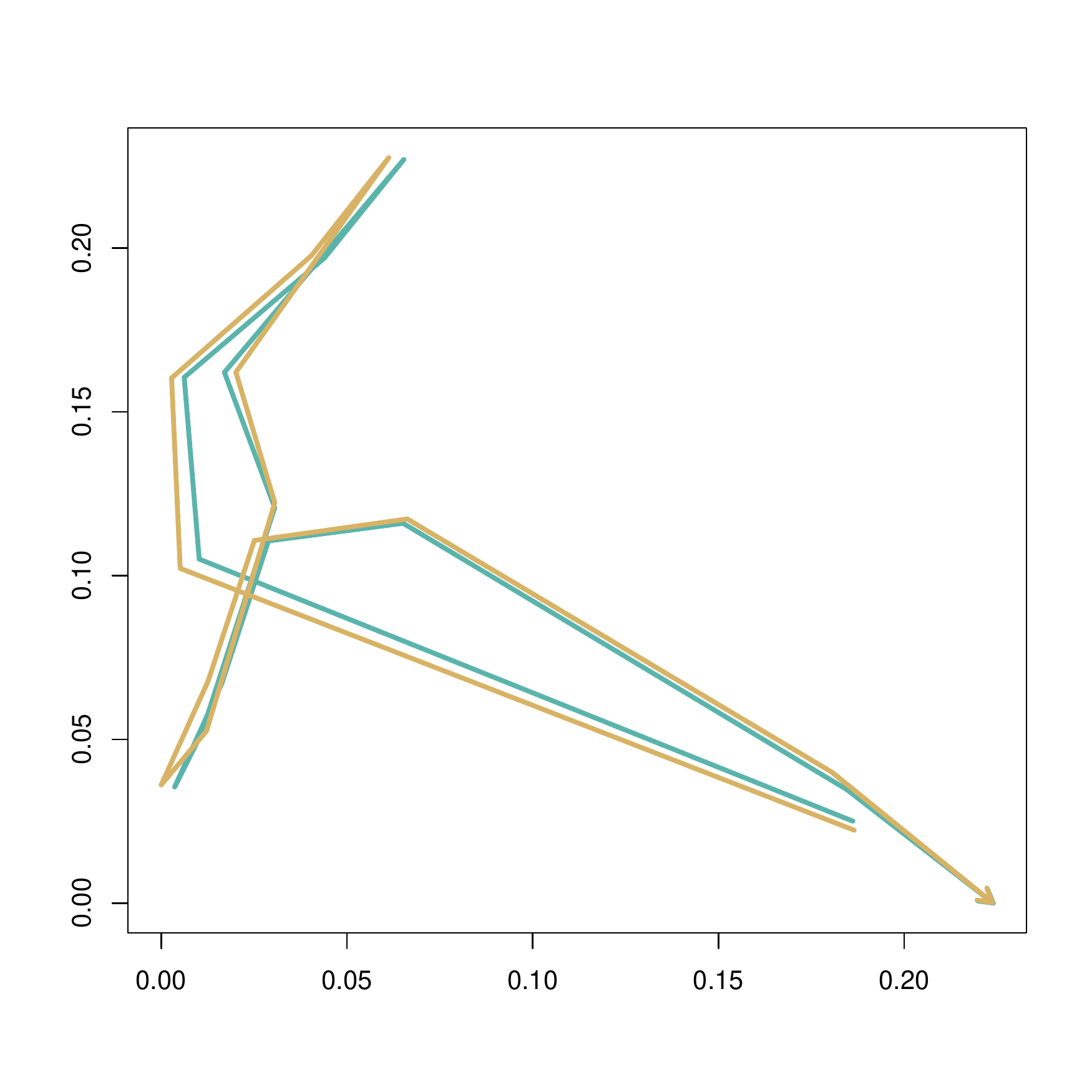}
		\subcaption{Dyad from cluster 1. \\ $Prox=1$; $DI_\theta = 1$; $DI_d = 0.98$}
	\end{minipage}
	\hfill				
	\begin{minipage}[b]{.5\linewidth}
			\centering	\includegraphics[scale=0.4]{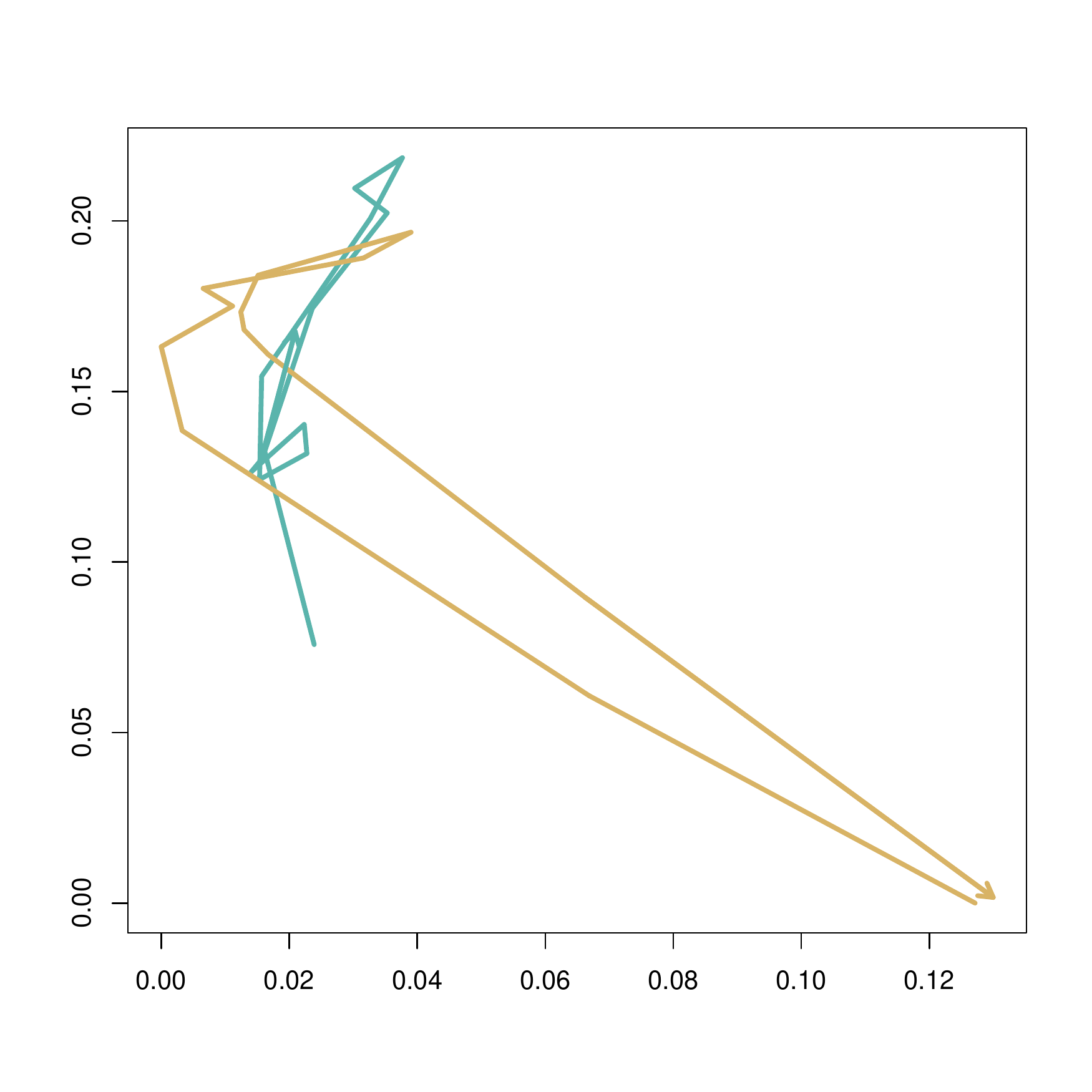}
			\subcaption{Dyad from cluster 2. \\ $Prox=0.57$; $DI_\theta = 0$; $DI_d = 0.69$}
		\end{minipage} \\
		\begin{minipage}[b]{\linewidth}
			\centering	\includegraphics[scale=0.4]{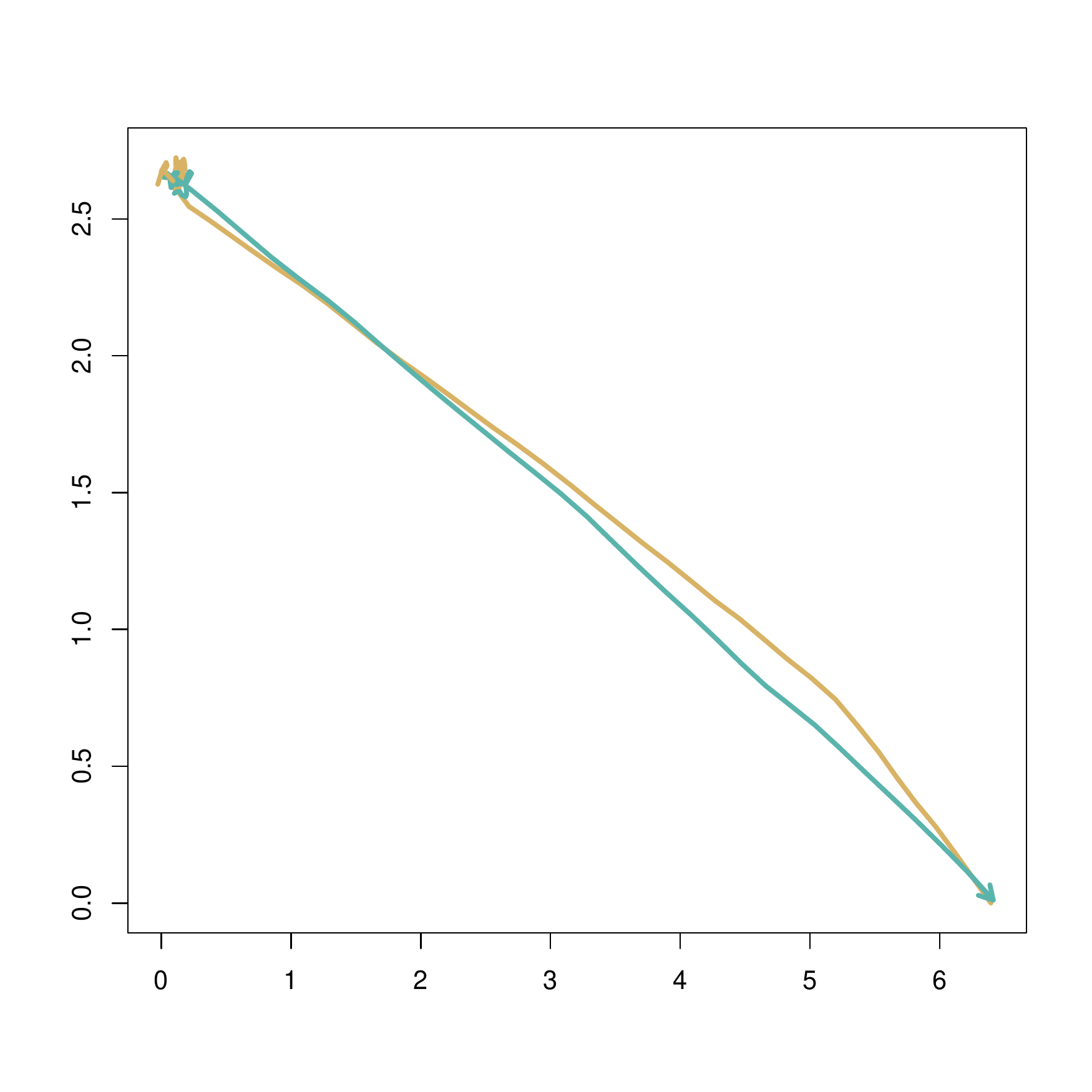}
			\subcaption{Dyad from cluster 3. \\ $Prox=0.06$; $DI_\theta = -0.07$; $DI_d = 0.24$}
		\end{minipage}
			\caption{The most representative dyadic example of each cluster for the pelagic pair trawler fleet, with the values of the metrics. The coordinates were transformed to avoid disclosing information about the vessels, whose identifiers are not shown either.}
	\label{Fig:PTMeg}				
\end{figure}

The three clusters obtained corresponded to distinct levels of joint movement (Fig. \ref{Fig:Clusters}). The first one (purple in Fig. \ref{Fig:Clusters}) corresponded to high joint movement in its three dimensions: proximity, coordination in direction and in speed/displacement. This was the expected pattern for partnership at sea. The second one (green in Fig. \ref{Fig:Clusters}) was associated to a lower degree of joint movement in all dimensions. The third cluster (yellow in Fig. \ref{Fig:Clusters}) was overall characterised by low proximity, relatively low coordination in direction, and low coordination in displacement. In these two metrics, there was a considerable amount of overlap, with Prox being the metric that made these two groups distinguishable. The tracks of the most representative dyad of each cluster, i.e. the one with the largest $P(Z=g \mid X=x)$, are shown in Fig. \ref{Fig:PTMeg}. Animations of the trajectories and time series related to the three metrics can be found in \url{https://rociojoo.github.io/partners-at-sea/}.

In total, $8\%$, $26\%$ and $66\%$ of the examined dyads were classified in the first, second and third cluster, respectively (Table \ref{Table:Dyads}). 
A low percentage of partners-at-sea dyads is not a surprise. 
The examined dyads were couples of vessel tracks coinciding in a common area at the same time. Not all pairs of vessels that cross their paths should be necessarily working together.
On the other hand, most of the vessels of the fleet, $56$ ($95\%$), participated at least once in dyads classified as partners at sea. From them, $46$ had exclusive partners (Fig. \ref{Fig:Networks}), which translated into a $0.82$ loyalty index for the fleet.

\subsection{Dyads from other fleets}

In this section, we focused only on the first group, i.e. partners at sea. 
The proportion of dyads classified in each cluster is presented in Table \ref{Table:Dyads}, and examples of dyads in each cluster for all fleets can be found in \url{https://rociojoo.github.io/partners-at-sea/}, a companion website for the manuscript.

When using $GMM_{pair trawlers}$ to classify dyads from the other fleets, we found partners at sea in all of them except for tuna purse seiners. In all the fleets, the posterior probabilities computed for classification were relatively high (medians were $>0.65$ and all posteriors were $>0.5$; Fig. \ref{Fig:proba}) showing low ambiguity for classification in all groups. 

For large, small bottom, mid-water otter trawlers and anchovy purse-seiners, $312$, $93$, $3$ and $568$ dyads were classified as partners at sea, respectively (Table \ref{Table:Dyads}). 
In all cases, it represented less than $1\%$ of the examined dyads, showing that vessels in the same area do not always move together, and when they do, they do not do it in large groups. 

\begin{figure}[ht!]	
	\begin{minipage}[b]{.3\linewidth}
		\centering
		\includegraphics[scale=0.4]{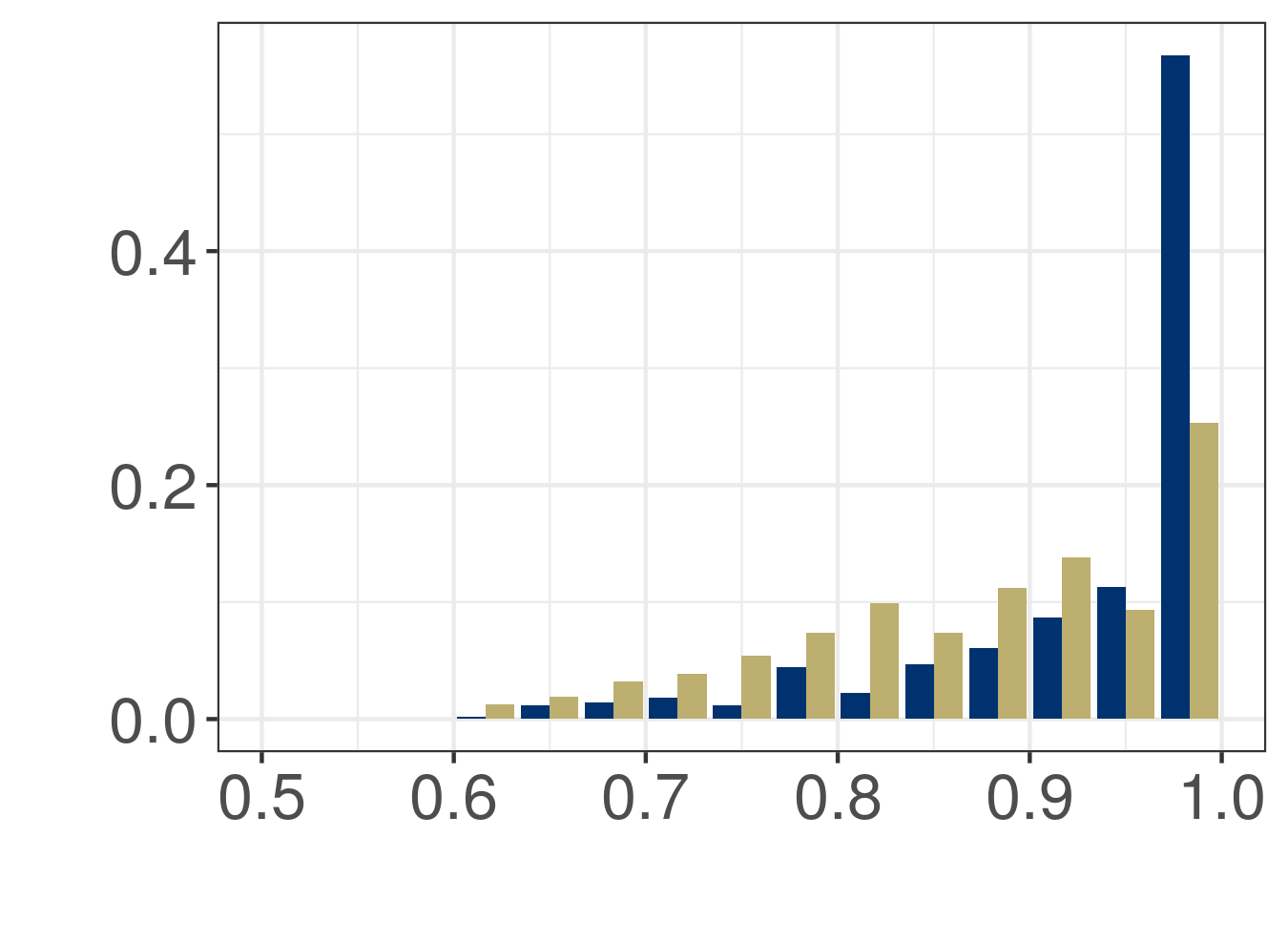}
			\subcaption{Large bottom otter trawlers - Prox}
	\end{minipage}		
	\hfill
	\begin{minipage}[b]{.3\linewidth}
		\centering
		\includegraphics[scale=0.4]{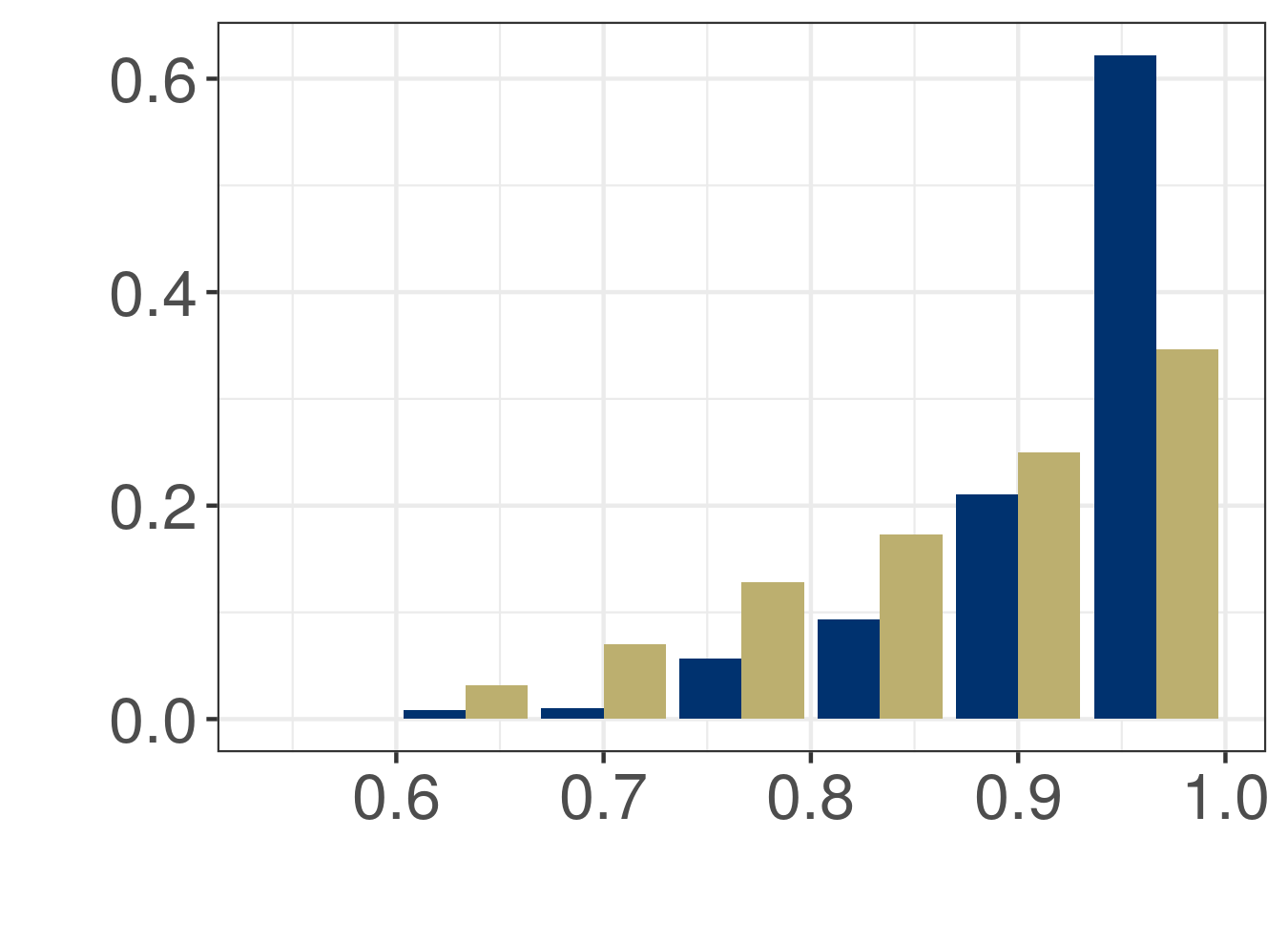}
				\subcaption{Large bottom otter trawlers - $DI_{\theta}$}
	\end{minipage}	
	\hfill
	\begin{minipage}[b]{.3\linewidth}
		\centering
		\includegraphics[scale=0.4]{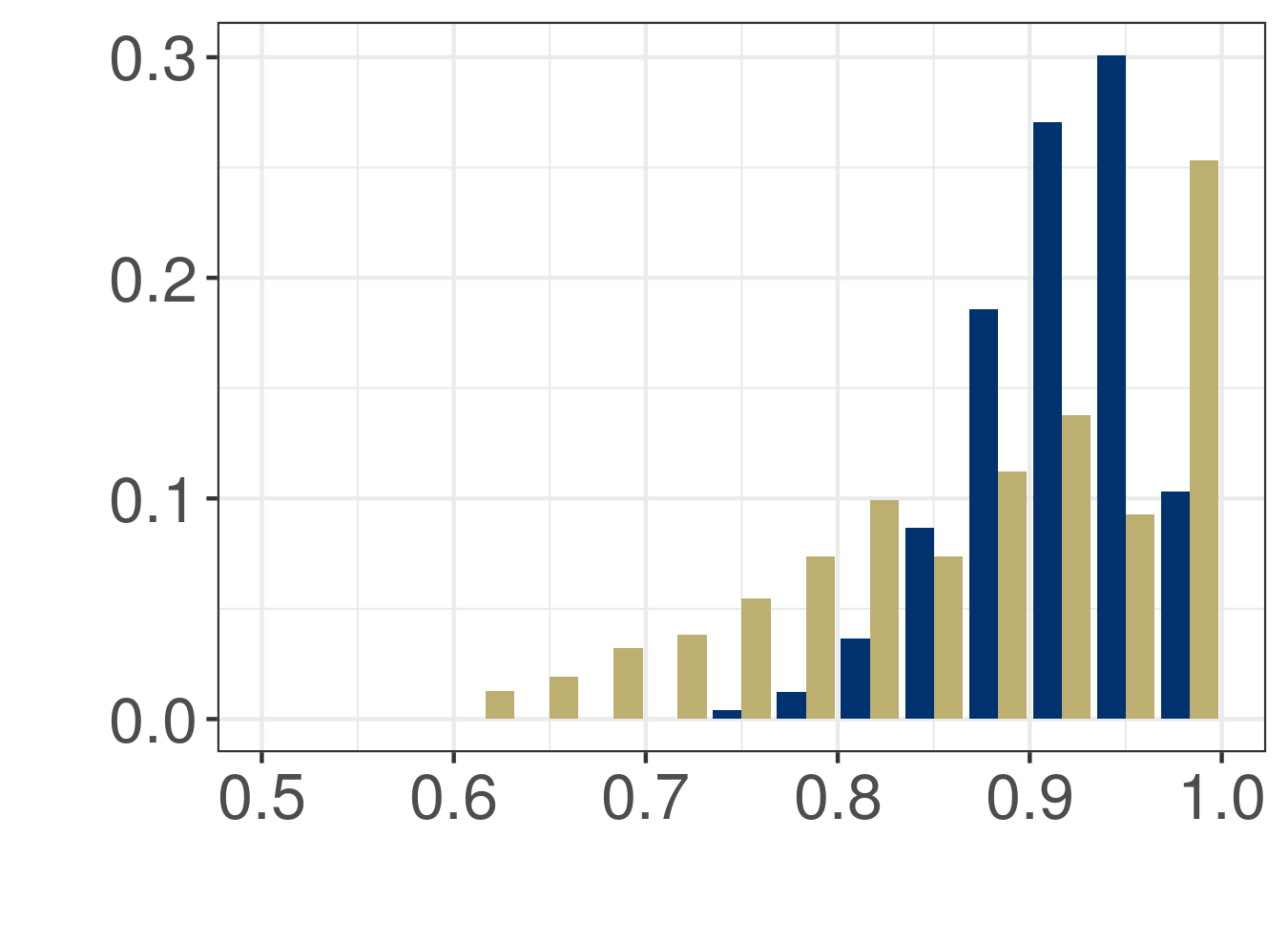}
				\subcaption{Large bottom otter trawlers - $DI_d$}
	\end{minipage}	
	
	\begin{minipage}[b]{.3\linewidth}
		\centering
		\includegraphics[scale=0.4]{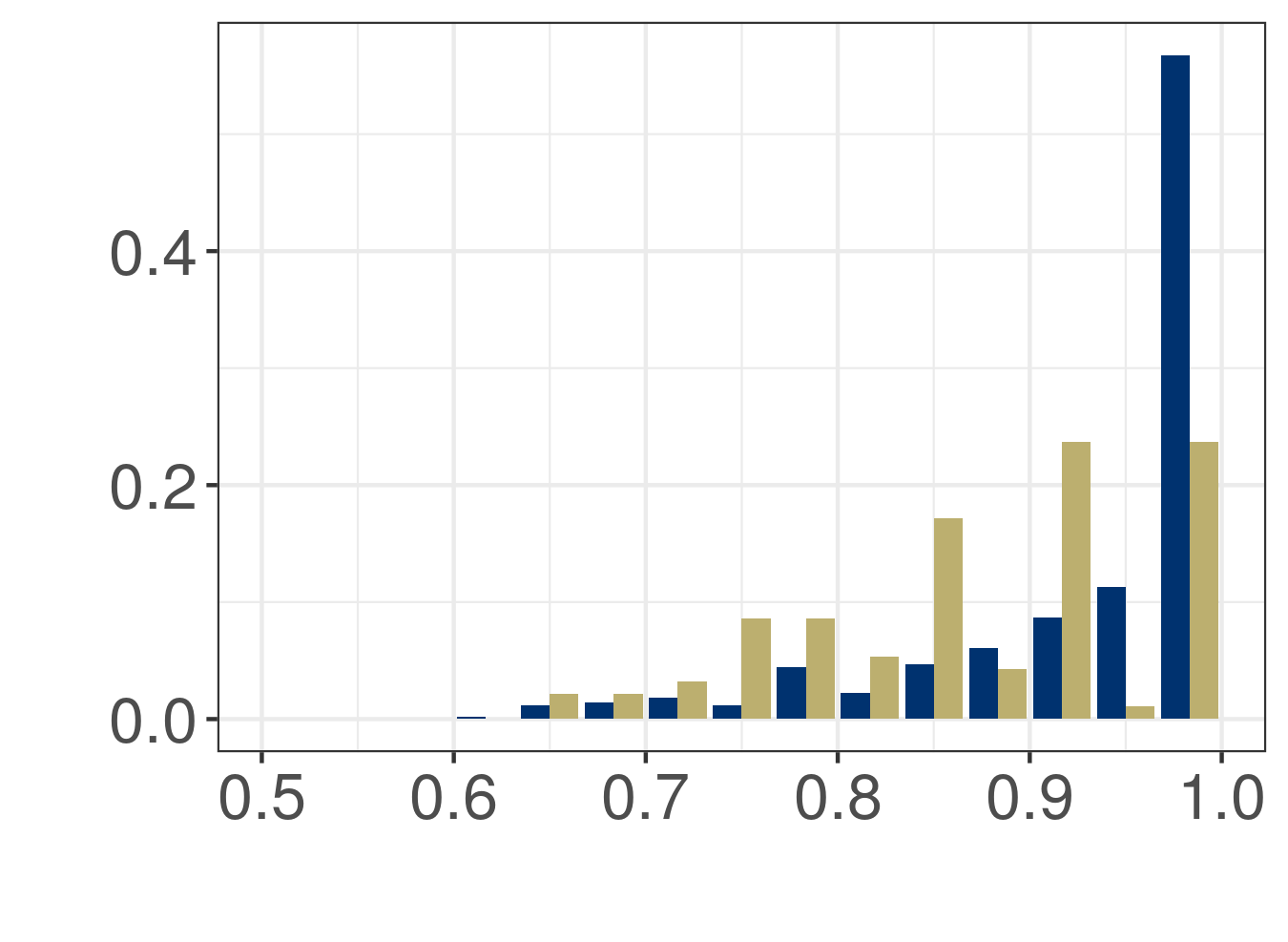}
			\subcaption{Small bottom otter trawlers - Prox}
	\end{minipage}		
	\hfill
	\begin{minipage}[b]{.3\linewidth}
		\centering
		\includegraphics[scale=0.4]{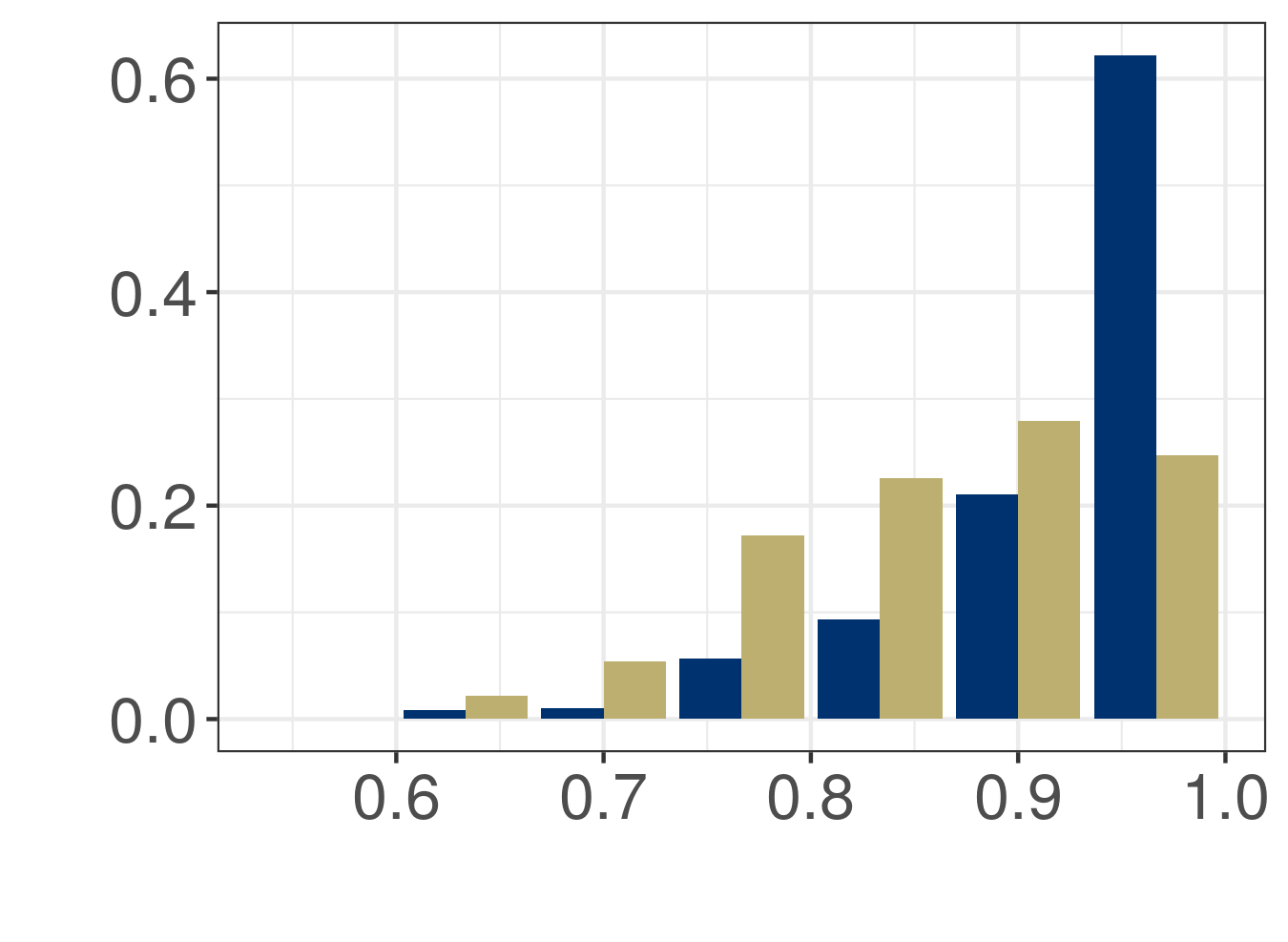}
				\subcaption{Small bottom otter trawlers - $DI_{\theta}$}
	\end{minipage}	
	\hfill
	\begin{minipage}[b]{.3\linewidth}
		\centering
		\includegraphics[scale=0.4]{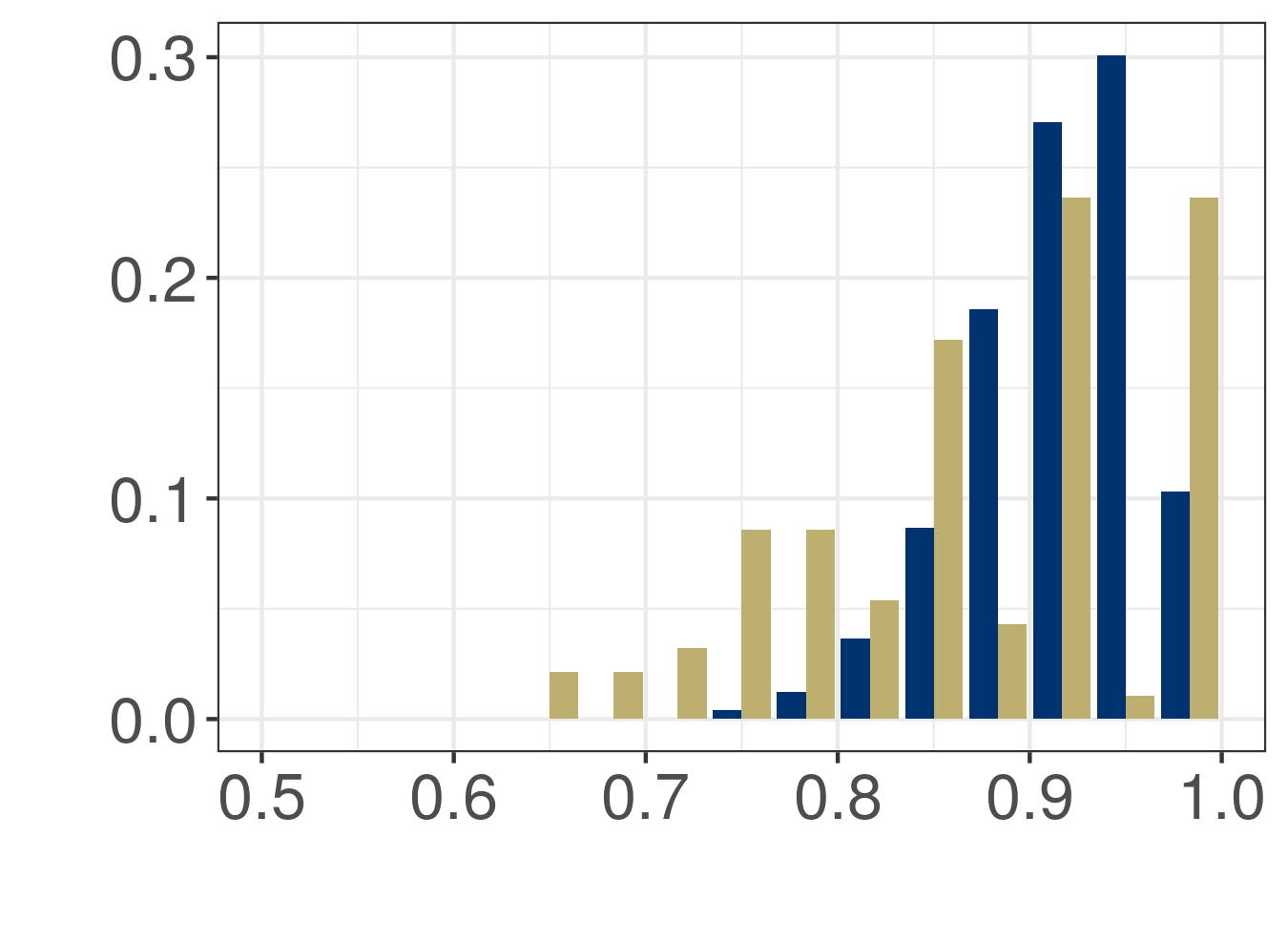}
				\subcaption{Small bottom otter trawlers - $DI_d$}
	\end{minipage}	
	
	\begin{minipage}[b]{.3\linewidth}
		\centering
		\includegraphics[scale=0.4]{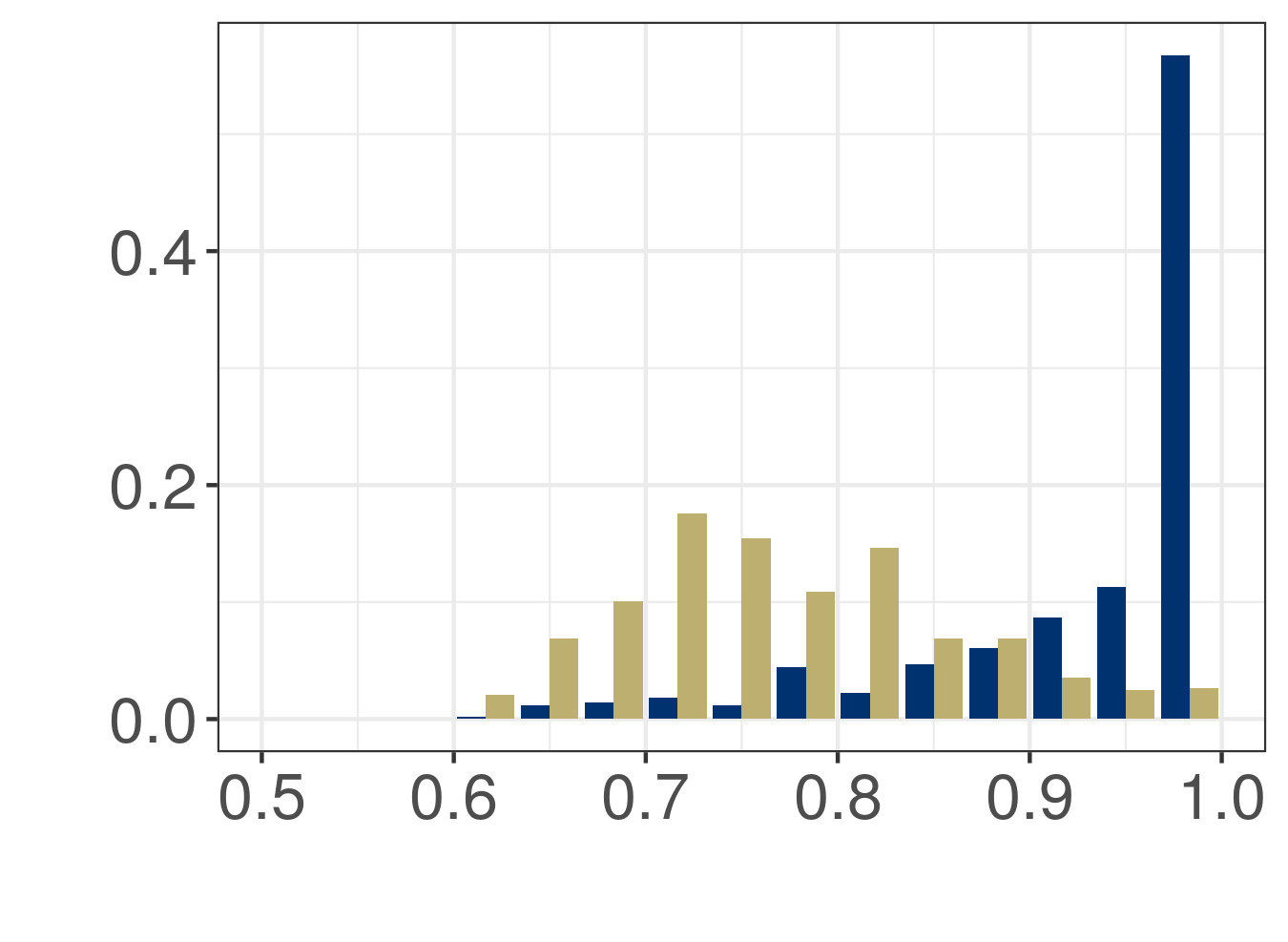}
			\subcaption{Anchovy purse-seiners - Prox}
	\end{minipage}		
	\hfill
	\begin{minipage}[b]{.3\linewidth}
		\centering
		\includegraphics[scale=0.4]{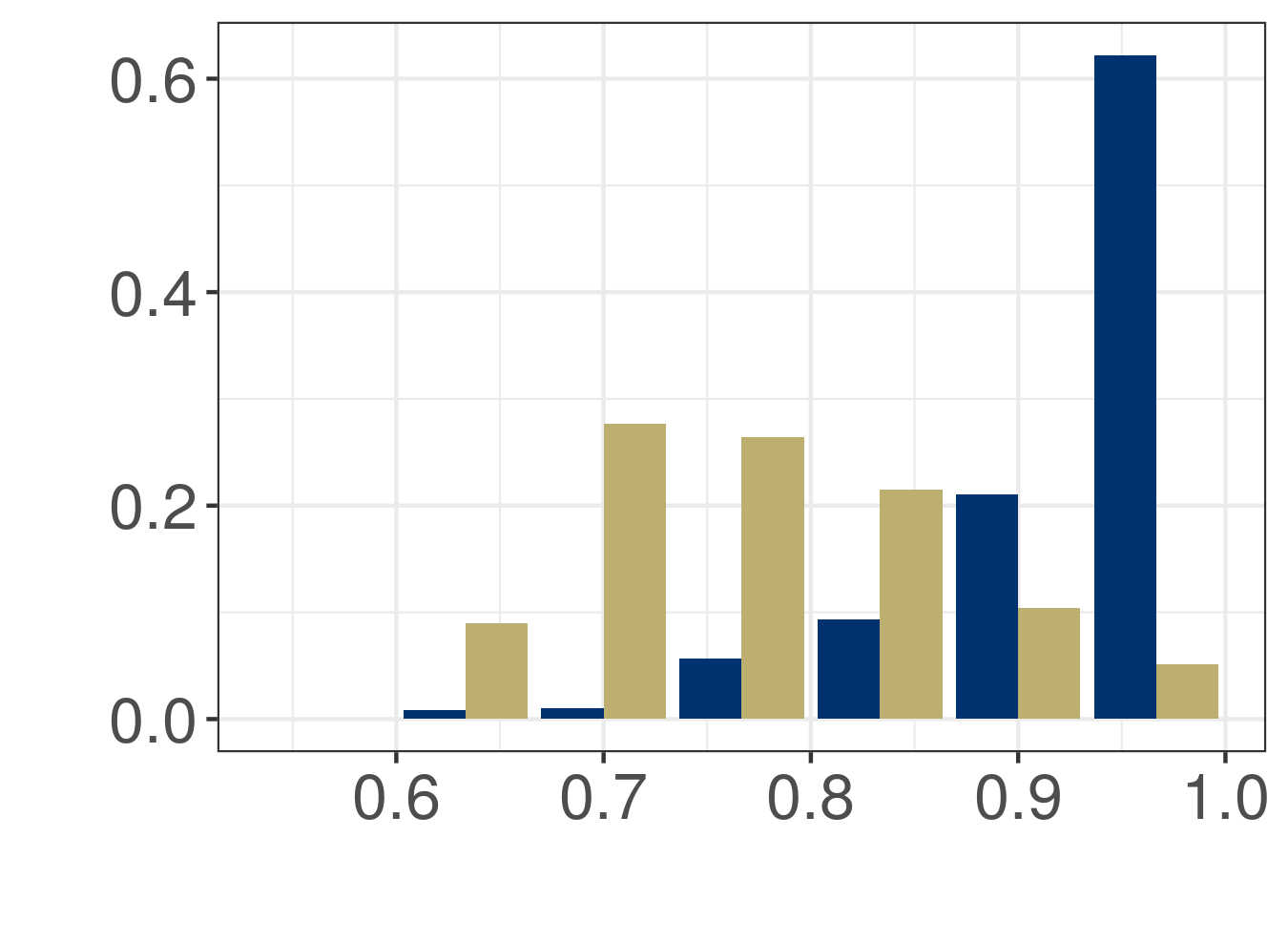}
				\subcaption{Anchovy purse-seiners $DI_{\theta}$}
	\end{minipage}	
	\hfill
	\begin{minipage}[b]{.3\linewidth}
		\centering
		\includegraphics[scale=0.4]{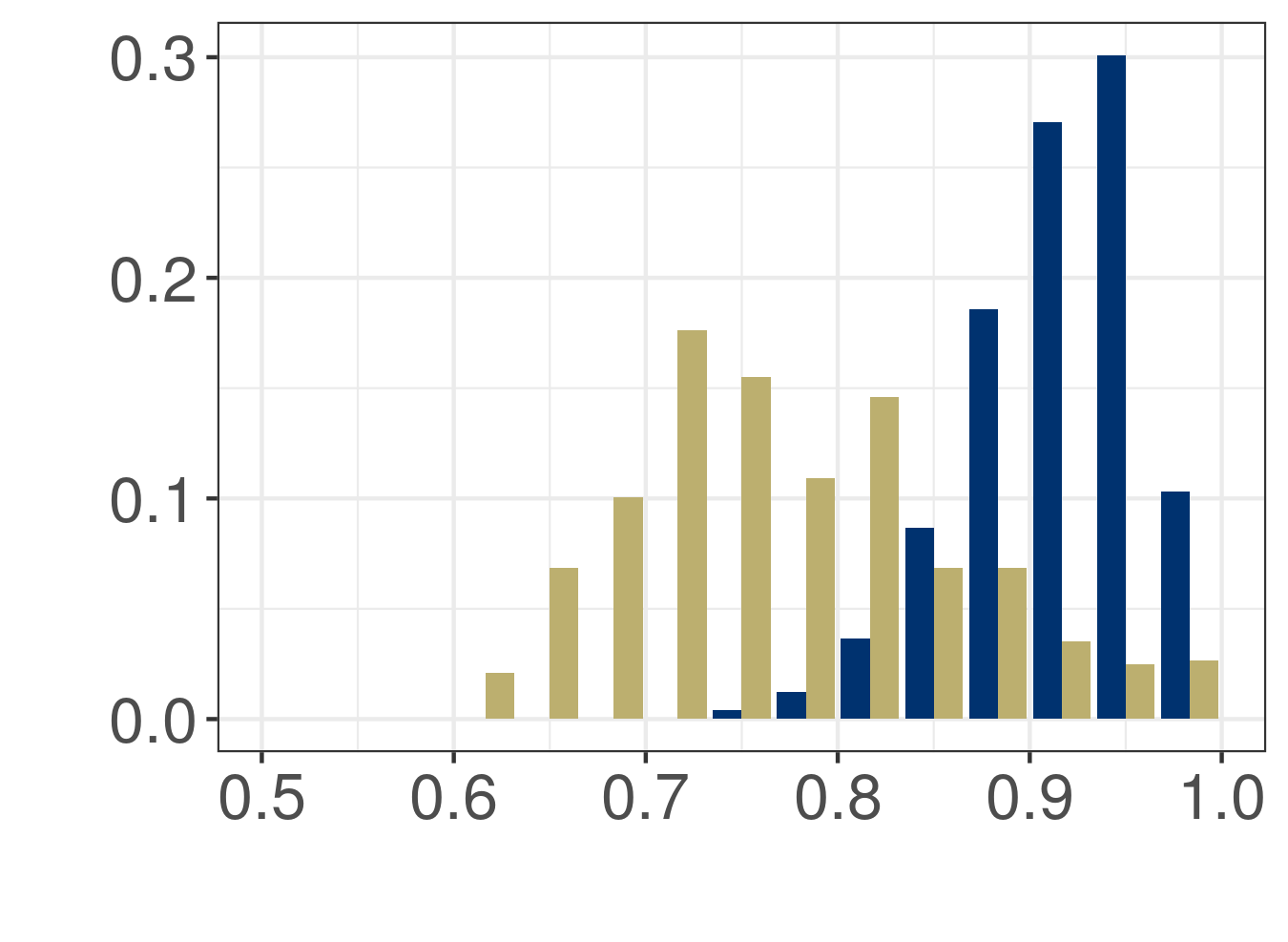}
				\subcaption{Anchovy purse-seiners - $DI_d$}
	\end{minipage}	
	\caption{Histograms of the joint movement metrics ($Prox$, $DI_{\theta}$, and $DI_d$, in the left, centre and right columns, respectively) for the first group or `partners at sea', comparing the pelagic pair trawlers (blue) with each of the other fleets (mustard). The other fleets are, in row order from top to bottom: large bottom otter trawlers, small bottom otter trawlers and anchovy purse-seiners. Tuna purse-seiners and mid-water otter trawlers  are not shown as no dyad and only three dyads, respectively, were associated with partnership.}
	\label{Fig:Hist1}
\end{figure}

We compared the distribution of values of the metrics in the first group between pelagic pair trawlers and the other fleets (large and small bottom otter trawlers, and anchovy purse-seiners; Fig. \ref{Fig:Hist1}).
Large bottom otter trawlers showed the most similar shapes of the distributions to pair trawlers, for all metrics, though the values of $DI_{d}$ were less skewed to the right than for pair trawlers. 
This difference in skewness for $DI_{d}$ was also true for the other two fleets. 
Moreover, `partners at sea' among anchovy purse-seiners took lower values of all the metrics (more skewed to the left). Since both fleets target pelagic species, one might have expected to find similar metric values for their partners at sea. This difference could be related to the different sampling rate (10 minutes), which allows looking at a finer scale in joint movement, showing that at this scale it is rather low. It could also be an indicator of a joint movement that does not occur at a dyadic scale, i.e. a couple of vessels that decide to move together; if larger groups were moving together, this pattern would not have necessarily reflected in very high values in the dyadic movement metrics. 

\begin{figure}[ht!]

 	\begin{minipage}[b]{.5\linewidth}
 		\centering%
 		\includegraphics[scale=0.15]{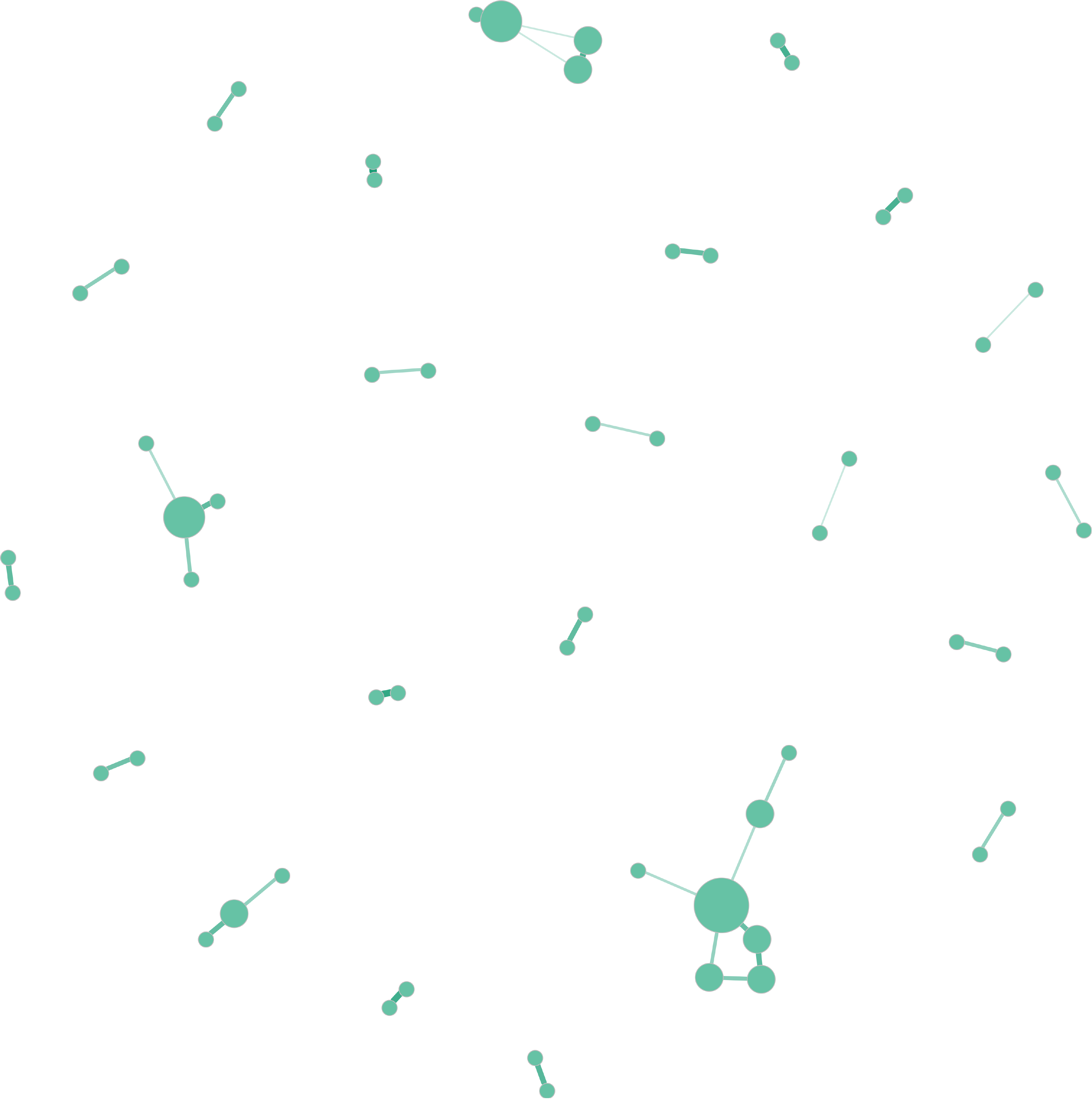}
 		\subcaption{Pelagic pair trawlers}
 		\label{PTMnet}
 	\end{minipage}
 	\hfill
 	\begin{minipage}[b]{.5\linewidth}
 		\centering%
 		\includegraphics[scale=0.15]{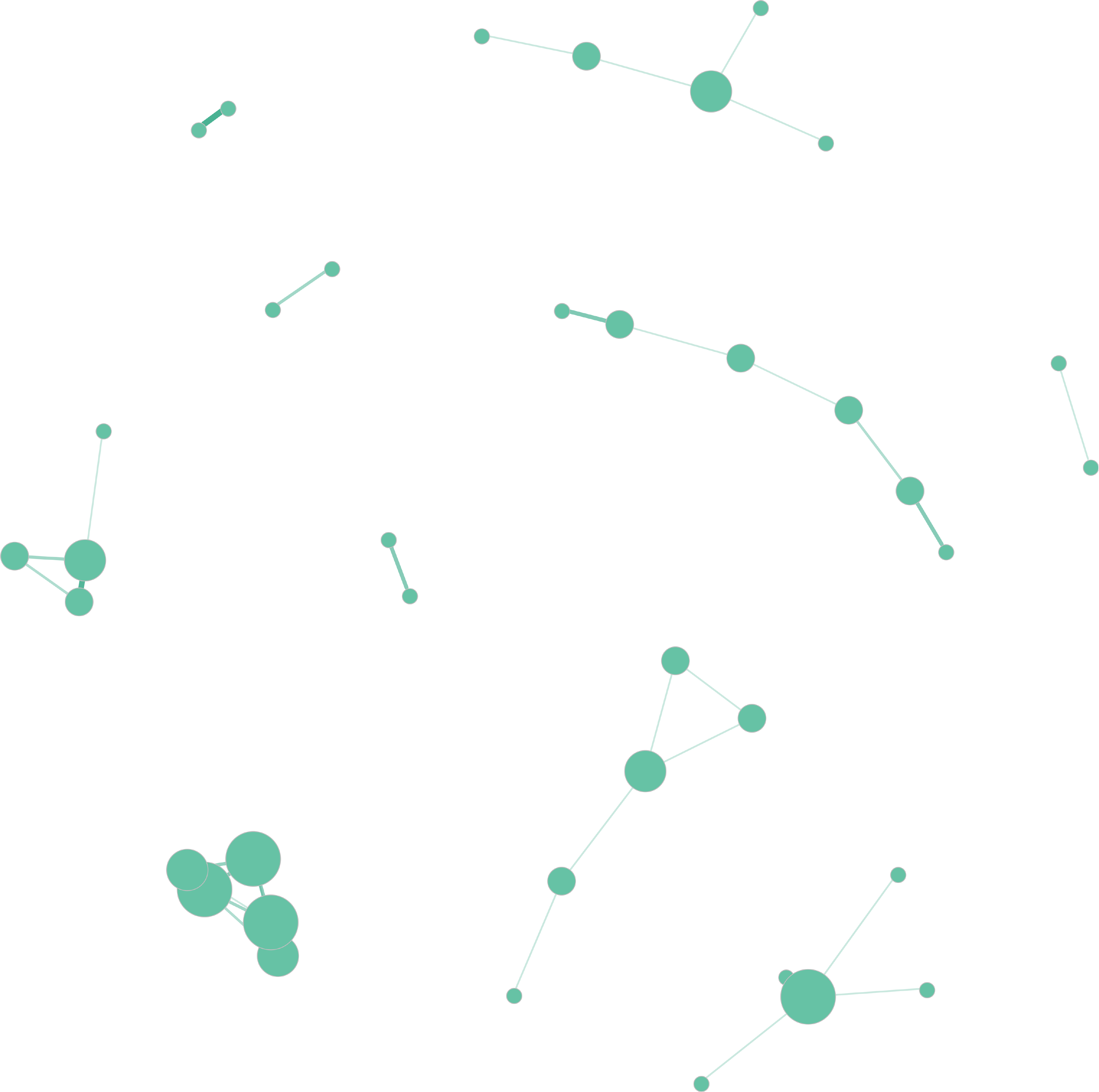}
 		\subcaption{Large bottom otter trawlers}
 		\label{OTBlargenet}
 	\end{minipage}
 	\begin{minipage}[b]{.5\linewidth}
 		\centering%
 		\includegraphics[scale=0.15]{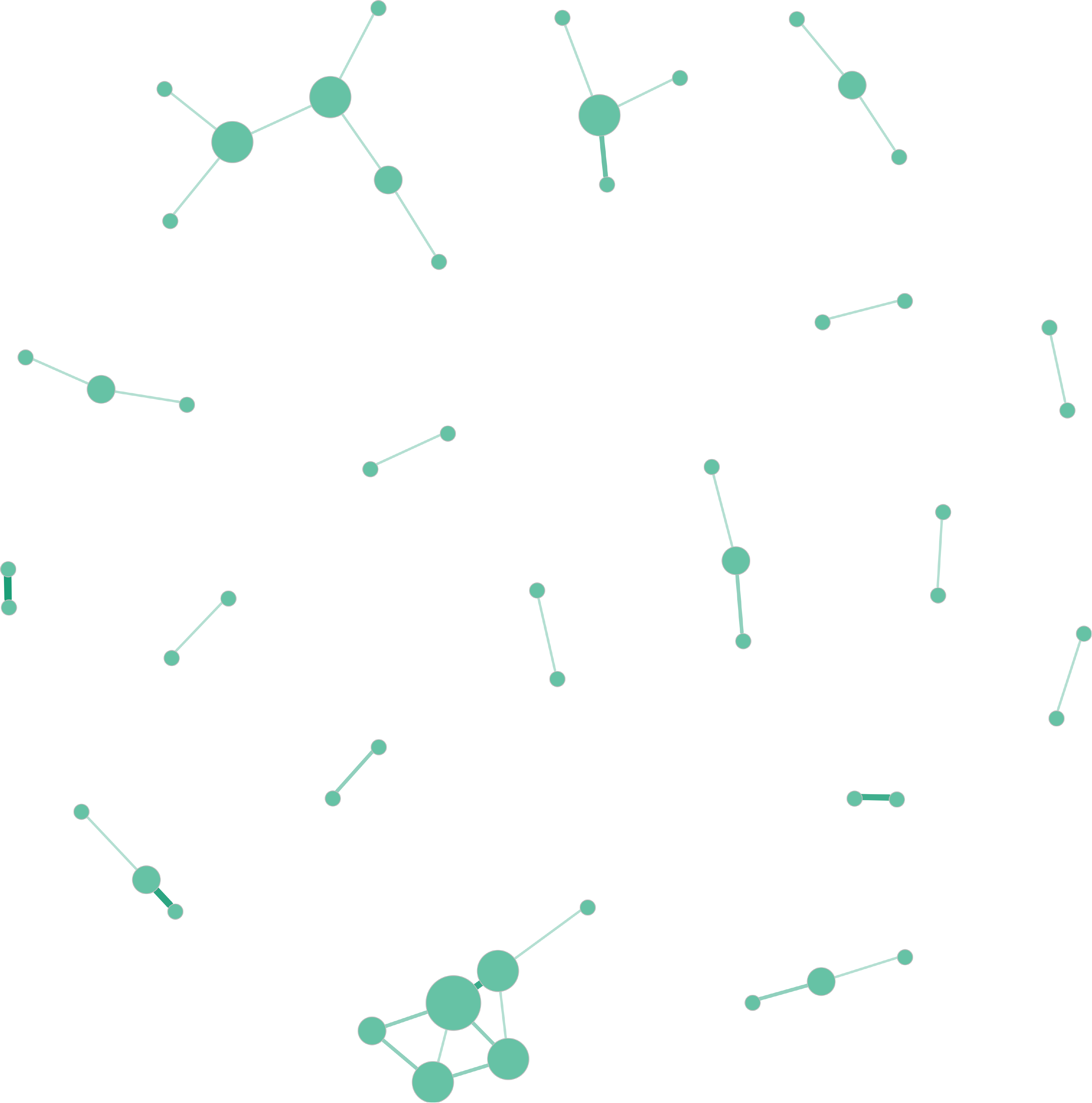}
 		\subcaption{Small bottom otter trawlers}
 		\label{OTBsmallnet}
 	\end{minipage}
 	\hfill
 	\begin{minipage}[b]{.5\linewidth}
 		\centering%
 		\includegraphics[scale=0.15]{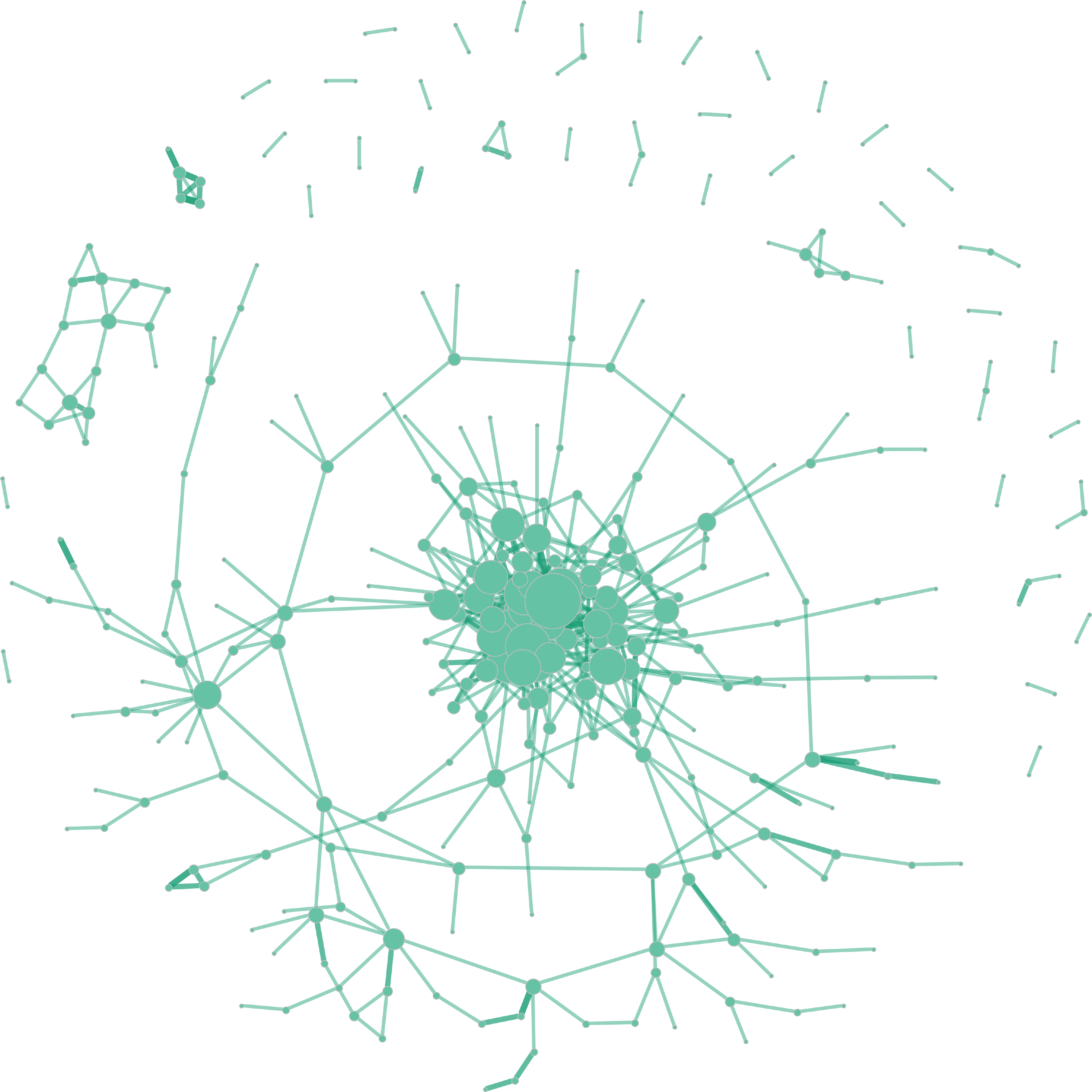}
 		\subcaption{Anchovy purse-seiners}
 		\label{APS10net}
 	\end{minipage}			
 	\caption{Network representation of partnership for the pelagic pair trawlers (a), small bottom otter trawlers (b), large bottom otter trawlers (c) and anchovy purse-seiners (d). Tuna purse-seiners and mid-water otter trawlers are not shown as no dyad and only three dyads, respectively, were associated with partnership. Within each network, only vessels that engaged in partnership at sea at least once were represented. The size of the nodes (vessels) are proportional to the number of times they were involved in partnership. The thickness of the lines between nodes are proportional to the number of partnerships between both nodes.}
 	\label{Fig:Networks}
 \end{figure}

The percentage of vessels engaged in at-sea partnership and their exclusiveness varied greatly among fleets (Fig. \ref{Fig:Networks}). $38$ out of $266$ large bottom otter trawlers ($14\%$) showed at-sea partnership at least once, and from them, $19$ had exclusive partners (loyalty = $0.50$). A larger percentage of small bottom otter trawlers engaged in partnership ($26\%$, or $52$ out of $202$). From them, $38$ had exclusive partners (loyalty = $0.73$). Only $4$ out of $70$ mid-water otter trawlers engaged in partnership, which was exclusive (loyalty = $1$) and only occurred three times. In contrast, $43\%$ of the anchovy purse-seiners engaged in partnership (or $327$ out of $757$ vessels). $134$ of these vessels were exclusive (loyalty = $0.41$). Most anchovy purse-seiners showed joint-movement links with large groups of vessels (Fig. \ref{Fig:Networks}d), which would be consistent with the differences in the metrics distribution (Fig. \ref{Fig:Hist1}).

 \section{Discussion}
 
In this work, we aimed at identifying partners at sea in different fleets around the world. We presented a simple heuristic approach to identify them by means of joint movement metrics \cite[]{Joo2018}, use of Gaussian mixture modelling, and taking pelagic pair trawlers as a `training' dataset. 

Partners at sea were identified in all the examined fisheries, except for tuna purse-seiners. 
This could be partly explained by the long duration of their fishing trips and large range of movement. While the trip duration in the other fleets ranged between less than a day and four days, tuna purse-seiner fishing trips lasted about 30 or 40 days.
Tuna purse-seiners, not bounded to fish together, showed that there was no strategy involving dyadic joint movement throughout their whole trips.
However, data exploration showed that some vessels moved together in pairs for parts of their trips (see \url{https://rociojoo.github.io/partners-at-sea/} for an example in group 2).
The identification of trip segments associated to joint movement (i.e. redefining a dyad) was out of the scope of this work, and remains open for future research. 

Mid-water and small bottom otter trawlers performed equally in terms of trip duration and distances covered. However, the mid-water otter trawler dataset only contained three partners-at-sea dyads, suggesting that individual competition could be higher in this fleet, or that working together would bring them no benefit, which could be due to their smaller fishing zones or the spatial behaviour of their targeted fish. 
Compared to mid-water trawlers, a higher percentage of both small and large bottom otter trawlers participated in partnerships, showing that this is a strategy used in these fleets, though it has not been adopted by the majority of the vessels.
These three trawler fleets are composed of vessels that engage in fishing activities (métiers) that target demersal or benthic species (fish, crustaceans, cephalopods). These métiers are likely to require less synchronous collaboration than pelagic métiers. Instead, the observed partner-at-sea behaviours could have been shaped by environmental or physical constraints \cite[e.g. currents,][]{Gloaguen2016} that the vessels would be facing in the same fishing area at the same time, rather than a collaborative fishing strategy. 

A third of anchovy purse-seiners moved in partnership at least once during the analysed fishing season. Though the trips had a short duration ($\sim 17$ hours), the sampling rate from these VMS data was very high ($\sim 10$ minutes).
At such resolution, joint movement patterns were identified. 
In this intensive and highly dynamic monospecific fishery, these findings are somehow a surprise that may be worth studying in more detail in the future.
The high number of vessels in this fleet showing joint movement, and the high number of connections displayed in its social network, makes it appealing to study joint movement in larger groups for this fleet. 

While it was expected to find partnership in pelagic pair trawlers, the degree of loyalty in this fleet was previously unknown, thus revealing about their partnership strategies. $82\%$ of the vessels (or fishers) opted for exclusive partnerships, and the ones who did not, exchanged partners in very reduced groups. In large and small bottom otter trawlers, the loyalty between vessels involved in the partner at sea cluster was lower; small bottom otter trawlers are involved in larger groups (Fig. \ref{Fig:Networks}). 
Non-exclusive partnerships involved even larger groups in the anchovy purse-seine fleet. 
These fleets may be revealing two opposed partnership strategies: exclusiveness, which would involve commitment or long-term partnership, and opportunism, in which a vessel would move jointly with another one (or even a group of vessels) without any previous history or commitment. We did not assess the associations between partnerships and belonging to a same company, and it could be appealing for future studies to analyze if this would correspond to a strategy where the ship-owner requires his fishing masters to work together. 

This work represents a first approach into studying joint movement behaviour and strategies in fisheries.
It highlights the fact that not all trajectories can be considered as independent, an assumption made in most modelling studies (e.g. using state space models; \cite{Joo2013,Gloaguen2015a}). Furthermore, it could be appealing to apply this approach to select, from a set of trajectories, those that do not show any partnership at sea. This could allow computing Catch per Unit of Effort only drawn from independent fishing operations. It could also be used to evaluate potential errors in modelling fleet dynamics. For instance, one could fit state-space models using independent tracks on one hand and using all the tracks on the other, and compare the goodness of fit of both models --and simulation results --to evaluate the biases in state estimations linked to the dependence between vessels. 

In this study, we focused on a very specific scale of joint movement, the dyad, defined as a unit composed of fishing trip segments of two vessels occurring at the same time and in a common area. 
Studying the strategies of fleets like the tuna purse-seiners could benefit from the development of methods to identify joint movement at smaller scales (e.g. segments of fishing trips).
In addition, for many fisheries, like the anchovy purse-seine fishery, the characterisation of joint movement in larger groups could help understanding the scales of collective behaviour in the fisheries.
Besides joint movement, leader/following dynamics would also be worth exploring (see a brief discussion in \cite{Joo2018}). 
All of these components would help characterising spatial behaviour patterns, but it would not be enough to understand the triggers of these behaviours. 
A next step would be to understand the associations between joint movement (or following movement) and external factors such as the spatial aggregation of the targeted species, the direction of currents, or management and economic policies. 
Ultimately, understanding and modelling fisher movement including its collective components will contribute to better estimations of local exploitation of resources. More realistic movement models would allow better simulations of fisher spatial behaviour and effort for different management scenarios, thus improving decisions for management.

\section{Acknowledgements}

The authors would like to thank Youen Vermard and Fabien Forget for useful feedback on the French fleets operating in the North-East Atlantic Ocean and the Indian Ocean, respectively. Youen's feedback on data processing was of great help, as well. We also acknowledge the collaboration of Ob7 – Observatoire des Ecosyst\`emes P\'elagiques Tropicaux exploit\'es, for the tuna dataset. The tuna data used in this study were collected through the Data Collection Framework (Reg 2017/1004 and 2016/1251) funded by both IRD and the European Union.
We are also grateful to Emily Walker for codes related to the tuna purse-seine fishery. Guidance for use of servers and different computers from Olivier Berthele and Audric Vigier were key in the first stages of this work, when Rocío Joo's work computer was a mess. Thanks to both of you.

\section{Authors' contributions}
RJ, SM and NB conceived the study. NG gave valuable insights on fishing behaviour at sea that were key to the study design and interpretation of results. RJ led the data processing and analysis, with contributions from PM and JR. MPE suggested and helped implementing the GMM. RJ led the writing of the manuscript. SM, NB and MPE made major contributions to the manuscript, and NG and PM made minor contributions to it. 

\section{Data availability statement}

The dyads' metrics along with all of the R codes for GMM and computation of the fleet characteristics are available on Zenodo: \url{https://doi.org/10.5281/zenodo.4016377}. The codes can also be viewed from \url{https://rociojoo.github.io/partners-at-sea/data-processing-and-analysis.html}. Due to confidentiality agreements, the raw VMS data cannot be shared. 

%

\newpage

\bibliographystyle{abbrv}

\end{document}